\documentclass[manuscript=article, layout=twocolumn]{achemso}
\setkeys{acs}{articletitle = true}

\usepackage[T1]{fontenc}
\usepackage[utf8]{inputenc}
\usepackage{graphicx}
\usepackage{amsmath}
\usepackage{color}
\usepackage{comment}
\usepackage{hyperref}
\usepackage{multirow}
\usepackage{subfig}
\usepackage{bm}
\usepackage{braket}
\usepackage[normalem]{ulem}

\title{Variational quantum eigensolver boosted by adiabatic connection}

\author{Mikul\'{a}\v{s} Matou\v{s}ek}
\altaffiliation{Faculty of Mathematics and Physics, Charles University, Prague, Czech Republic}
\affiliation{J. Heyrovsk\'{y} Institute of Physical Chemistry, Academy of Sciences of the Czech \mbox{Republic, v.v.i.}, Dolej\v{s}kova 3, 18223 Prague 8, Czech Republic}

\author{Katarzyna Pernal}
\affiliation{Institute of Physics, Lodz University of Technology, \mbox{ul.\ Wolczanska 217/221, 93-005 Lodz, Poland}}
\email{pernalk@gmail.com}

\author{Fabijan Pavo\v{s}evi\'{c}}
\affiliation{Algorithmiq Ltd., Kanavakatu 3C, FI-00160 Helsinki, Finland}
\email{fpavosevic@gmail.com}

\author{Libor Veis}
\email{libor.veis@jh-inst.cas.cz}
\affiliation{J. Heyrovsk\'{y} Institute of Physical Chemistry, Academy of Sciences of the Czech \mbox{Republic, v.v.i.}, Dolej\v{s}kova 3, 18223 Prague 8, Czech Republic}


\begin{document}

\begin{abstract}
In this work we integrate the variational quantum eigensolver (VQE) with the adiabatic connection (AC) method for efficient simulations of chemical problems on near-term quantum computers. Orbital optimized VQE methods are employed to capture the  strong correlation within an active space and classical AC corrections recover the dynamical correlation effects comprising electrons outside of the active space.
On two challenging strongly correlated problems, namely the dissociation of N$_2$ and the electronic structure of the tetramethyleneethane biradical, we show that the combined VQE-AC approach enhances the performance of VQE dramatically. Moreover, since the AC corrections do not bring any
additional requirements on quantum resources or measurements, they can literally boost the VQE algorithms. Our work paves the way
towards quantum simulations of real-life problems on near-term quantum computers.

\end{abstract}

\maketitle

\section{Introduction}

Quantum computers represent one of the most promising recent technological advances.
Despite complications inherently connected to storing and manipulating information in fragile quantum states of matter, quantum computing offers enormous computational power, with the potential to exponentially speed-up the solution of certain types of problems~\cite{nielsen_chuang}. Predicting the electronic structure of strongly correlated molecules and materials is considered as one such problem and a hot candidate for practical quantum supremacy~\cite{Cao2019,McArdle2020,HeadMarsden2020,Bauer2020,Motta2021}.

The first proposal of using quantum computers for molecular electronic structure calculations was put forward by Aspuru-Guzik et al.~\cite{aspuru-guzik_2005} who employed the Quantum Phase Estimation (QPE) algorithm~\cite{abrams_1997,abrams_1999,nielsen_chuang}. QPE is regarded as an ultimate quantum algorithm for finding eigenvalues of molecular Hamiltonians, which paves the way for very accurate simulations of the most complex molecular systems. However, due to very deep quantum circuits~\cite{Reiher2017}, its application on non-trivial problems requires a robust quantum error correction (QEC)~\cite{Terhal2015}, which is still out of reach of current and near-term future quantum devices.

Present-day and near-term quantum computers belong to the category of so called Noisy Intermediate-Scale Quantum (NISQ) devices~\cite{Preskill2018}. While still not offering the full QEC, NISQ devices are believed to offer important computational advantages and many NISQ-oriented quantum algorithms for simulations in chemistry, physics, and material science have been proposed in recent years~\cite{Cao2019,McArdle2020, HeadMarsden2020,Bauer2020,Motta2021}, most notably the Variational Quantum Eigensolver (VQE)~\cite{Peruzzo2014, McClean2016, Tilly2022}. VQE trades the long coherent quantum circuits of QPE for much shorter circuits with a large number of measurements and could offer a quantum advantage on NISQ devices in the near future, i.e. allow accurate simulations of classically intractable problems.

Nevertheless, despite the huge potential of NISQ algorithms, they certainly cannot treat larger molecular systems since mapping of the whole system onto a quantum register would be too demanding on quantum resources and eventually also require QEC. Instead, the concept of the complete active space (CAS)~\cite{Romero2018} can be employed and VQE or alternative NISQ algorithms~\cite{Smart2021, Stair2021} can be used to account for the strong correlation within the CAS. In fact, an efficient treatment of few dozens of strongly correlated electrons would in principle be sufficient to solve the most challenging strongly correlated problems of computational chemistry, such as the Fe-Mo cofactor~\cite{Reiher2017}. 

The efficient treatment of strong correlation via VQE must, however, be coupled with adequate treatment of the missing (out-of-CAS) dynamical electron correlation in order to achieve a chemical accuracy. The existing approaches can be classified into two categories. The first is based on the dimensionality reduction of the many-body Hamiltonians used in quantum algorithms which includes for example the theory of transcorrelated Hamiltonians or downfolding methods~\cite{Motta2020, Sokolov2023, Bauman2019, Bauman2021, Le2023}. The second category comprises corrections by classical post-processing methods such as quantum subspace expansion (QSE)~\cite{McClean2017, Takeshita2020}, or strongly contracted second-order $N$-electron valence perturbation theory (NEVPT2)~\cite{Krompiec2022}. Very recently, the combination of QSE and NEVPT2, which corresponds to the partially contracted NEVPT2 scheme, has been presented~\cite{Tammaro2023}. 
The methods from both the aforementioned categories suffer to some extent from inherent limitations. The former need to neglect higher-body terms in order to work with computationally tractable two-body Hamiltonians, and the latter 
methods can improve
dramatically upon CAS VQE solution, however, at the cost of additional VQE measurements, since their exact formulations require up to 4-electron active space reduced density matrices (RDMs).

Herein, we propose a novel method which belongs to the second category mentioned above and
integrate the VQE algorithm with the recently developed adiabatic connection (AC) methods for multireference wave functions~\cite{Pernal2018, Pastorczak2018, Pastorczak2019}.
Our approach avoids the limitation of higher-body RDMs, since AC methods require only up to 2-electron active space RDMs (2-RDMs), quantities directly available from the VQE procedure. Therefore neither additional quantum resources in terms of qubits or quantum gates, nor additional quantum measurements are needed compared to the VQE algorithm. In fact, the AC dynamical electron correlation corrections may be computed by means of classical computational methods with only a modest additional overhead~\cite{Drwal2022} and thus literally boost the VQE algorithm.

Recently, an alternative approach based on the multiconfiguration pair-density functional theory (PDFT), which similarly to AC requires only up to 2-RDMs, combined with the quantum solver of contracted eigenvalue equations has been presented~\cite{Boyn2021}. We would like to stress that our approach, in contrast to PDFT: (a) is strictly free of the double counting problem of the electron correlation, (b) converges to the full configuration interaction (FCI) with the expansion of active space, and (c) is free from approximate density functionals.

In what follows, we briefly review the basics of both, the VQE algorithm and the AC methodology. The combined VQE-AC approach is then applied and by means of classical numerical simulations tested on two challenging strongly correlated problems, namely the dissociation of nitrogen molecule (N$_2$) and the tetramethyleneethane (TME) biradical.

\section{Methods}

\subsection{Variational Quantum Eigensolver}

The VQE method~\cite{Peruzzo2014, McClean2016} combines classical variational energy minimization over normalized trial wave functions $\Psi(\vec{\theta})$ parametrized with $\vec{\theta}$

\begin{equation}
  E = \min_{\vec{\theta}} \bra{\Psi(\vec{\theta})} \hat{H} \ket{\Psi(\vec{\theta})}
\end{equation}

\noindent
with a state preparation and a measurement of the Hamiltonian expectation value on a quantum computer. This way, one can avoid deep circuits of QPE and replace them by shallow circuits corresponding to trial states preparation. Nevertheless, the crucial ability of an $n$-qubit quantum register to store a $2^n$-dimensional state vector is still exploited. 

After preparing the trial state (or ansatz) $\Psi(\vec{\theta})$, the expectation value of the Hamiltonian is measured. When working with the second-quantized representation of the Hamiltonian within the Born-Oppenheimer approximation (going beyond this approximation
is also possible~\cite{Veis2016,pavosevic2020chemrev,Pavosevic2021})

\begin{equation}
  \hat{H} = \sum_{pq} h_{pq} \hat{a}_{p}^{\dagger} \hat{a}_{q} + \frac{1}{2} \sum_{pqrs} \langle pq | rs \rangle \hat{a}_{p}^{\dagger} \hat{a}_{q}^{\dagger} \hat{a}_{s} \hat{a}_{r}
  \label{ham_sec_quant}
\end{equation}

\noindent
where $h_{pq}$ and $\langle pq | rs \rangle$ denote one and two-electron integrals in the molecular spin orbital basis~\cite{szabo_ostlund}, the actual energy can be computed by contraction of the integrals $h_{pq}$ and $\langle pq | rs \rangle$ with 1-RDM ($\gamma_{pq}$) and 2-RDM ($\Gamma_{pqrs}$)

\begin{eqnarray}
  E(\vec{\theta}) & = & \sum_{pq} h_{pq} \gamma_{pq}(\vec{\theta}) \nonumber \\
  & + & \frac{1}{2} \sum_{pqrs} \langle pq | rs \rangle \Gamma_{pqrs}(\vec{\theta}) \\
  \gamma_{pq}(\vec{\theta}) & = & \bra{\Psi(\vec{\theta})} \hat{a}_{p}^{\dagger} \hat{a}_{q} \ket{\Psi(\vec{\theta})} \label{g1} \\
  \Gamma_{pqrs}(\vec{\theta}) & = & \bra{\Psi(\vec{\theta})} \hat{a}_{p}^{\dagger} \hat{a}_{q}^{\dagger} \hat{a}_{s} \hat{a}_{r} \ket{\Psi(\vec{\theta})} \label{g2}
\end{eqnarray}

\noindent
The standard procedures of fermion-to-qubit mappings~\cite{Claudino2022}, such as the Jordan-Wigner~\cite{jordan_1928, whitfield_2010}, or Bravyi-Kitaev \cite{Bravyi2002, Seeley2012}, allow to represent the strings of second-quantized operators in (\ref{g1}) and (\ref{g2}) by combinations of products of Pauli operators (Pauli strings $\in \{ \sigma_x, \sigma_y, \sigma_z, I \}$) and their expectation values can be obtained by direct measurement on qubits corresponding to spin orbital indices (or alternatively by the Hadamard test).

Then computed energy is passed together with the actual values of $\vec{\theta}$ to a classical optimization routine. This can be a gradient-free optimization, such as Nelder-Mead simplex, or some gradient-descent, since strategies to directly measure the energy gradients were also developed \cite{McArdle2020}. The classical optimization produces the new set of parameters $\vec{\theta}$ and the whole procedure is repeated until energy convergence.

Different types of $\Psi(\vec{\theta})$ have been used in connection with VQE. The only condition, which must be fulfilled in order to preserve an efficiency of the VQE procedure is that the size of the parameter vector $\vec{\theta}$ scales polynomially with system size. 
The hardware efficient ansatz, which comprises the limited parametrized gate set, easy to implement on a given quantum architecture, is one example. These approaches do not use any information about a system studied and they have been successfully demonstrated on several small molecules~\cite{Kandala2017, Cao2019}. Another category comprises chemically inspired ansatze. The VQE methods of this type are almost exclusively based on the unitary coupled cluster theory (UCC)~\cite{Anand2022} and employ the following ansatz

\begin{eqnarray}
  \ket{\Psi(\vec{\theta})} & = & \hat{U}(\vec{\theta}) \ket{\Psi_{\text{ref}}} \\
  \hat{U}(\vec{\theta}) & = & e^{\hat{T}-\hat{T}^{\dagger}} \label{ucc_anz}
\end{eqnarray}

\noindent
where $\hat{U}(\vec{\theta})$ is a unitary operator, $\hat{T}$ represents the CC cluster operator and $\Psi_{\text{ref}}$ is an easy-to-prepare reference wave function, usually the Hartree-Fock (HF) Slater determinant. In case of UCCSD, $\hat{T}$ is restricted to particle-hole single and double excitations 

\begin{eqnarray}
  \hat{T} & = & \sum_{i,a} \theta_{i}^{a} \hat{a}^{\dagger}_a \hat{a}_i + \sum_{i<j,a<b} \theta_{ij}^{ab} \hat{a}^{\dagger}_a \hat{a}^{\dagger}_b \hat{a}_j \hat{a}_i \\
  & & i,j \in \text{occup.}; a,b \in \text{virt.} \nonumber
\end{eqnarray}

\noindent
and the CC amplitudes $\theta_{i}^{a}$ and $\theta_{ij}^{ab}$ form the parameter vector $\vec{\theta}$. Generalized formulations, which do not distinguish between occupied and virtual indices have also been developed and numerically tested in the context of quantum simulation~\cite{Lee2018, Bauman2021, Matsuzawa2020}.

In general, no convergent truncation for the UCCSD expectation value of energy is known, i.e. 
UCCSD method cannot be performed efficiently on a classical computer (due to infinite commutator expansions).
On the contrary, the unitary operation $\hat{U}(\vec{\theta})$ in (\ref{ucc_anz}) can be arbitrarily accurately approximated by a polynomial number of elementary single and two-qubit quantum gates~\cite{McClean2016}. In other words, UCCSD can be performed efficiently on a quantum computer by means of VQE. Moreover, the variational UCCSD method is expected to outperform traditional CCSD in accuracy,~\cite{culpitt2023unitary} especially 
for some strongly correlated problems like breaking of covalent bonds~\cite{McClean2016,Sokolov2020,Anand2022}.

Since $\hat{T}$ and $\hat{T}^{\dagger}$ in Eq.~\ref{ucc_anz} do not commute, the exponential of a summation of excitations cannot be written as a product of individual exponentials. In order to decompose $e^{\hat{T}-\hat{T}^{\dagger}}$ to elementary quantum gates, 
some sort of numerical approximation has to be used, e.g. the
first-order Suzuki-Trotter approach

\begin{equation}
  e^{A + B} = \left( e^{A/n} e^{B/n} \right)^n + \mathcal{O}(1/n)
\end{equation}

\noindent
In fact, even a single Trotter step ($n=1$) can provide the circuit ansatze, which yield accurate results, and have been used almost exclusively in UCCSD-VQE numerical studies~\cite{McArdle2020} (the so called disentangled UCCSD~\cite{Evangelista2019}). Part of the reason is that variational optimization absorbs most of the energy difference between the conventional UCCSD and the Trotterized form~\cite{McClean2016}. After the fermion-to-qubit mapping, the exponentials of excitation operators, such as $\exp{\theta_i (\hat{a}^{\dagger}_a \hat{a}_i - \hat{a}^{\dagger}_i \hat{a}_a})$, can be implemented with single-qubit rotations and CNOT gates \cite{whitfield_2010}.

As mentioned above, UCCSD is superior to CCSD and can accurately treat strongly correlated systems, nevertheless, due to its nature, it cannot fully approach FCI, i.e. an exact wave function. Moreover, its hardware implementation requires very deep quantum circuits. On the other hand, the so called Adaptive Derivative-Assembled Pseudo-Trotter ansatz Variational Quantum Eigensolver (ADAPT-VQE)~\cite{Grimsley2019} constitutes a UCC-type parametrization, which in contrast to naive UCCSD aims to adaptively build a parametrization that is able to approximate a FCI wavefunction with much more shallow quantum circuits. 
The ansatz is constructed in a way that operators from a given pool are sequentially added to the ansatz based on their contribution to the energy. More compact and more accurate quantum circuits can be achieved this way. ADAPT-VQE was later further improved in terms of circuit depth by using a coupled cluster-like ansatz which is constructed directly in the qubit representation~\cite{Ryabinkin2018} and named qubit-ADAPT-VQE~\cite{Tang2021}.

\subsection{Adiabatic Connection}

The AC theory for multireference wave functions~\cite{Pernal2018}  is a general approach to the correlation energy calculations, which can be applied to any reference wave function. If the wavefunction is of the complete active space (CAS) form, i.e.\ it is constructed from inactive (doubly occupied) and active (fractionally occupied) orbitals (the remaining orbitals form a set of virtual orbitals) then the AC approximations aim to recover the out-of-active-space correlation energy missing in the CAS model. The total electronic energy can be written as

\begin{equation}
 E = \bra{\Psi_{\text{CAS}}} \hat{H} \ket{\Psi_{\text{CAS}}} +E^{\text{AC}}_{\text{corr}}
\end{equation}

\noindent
where $\hat{H}$ is the exact Hamiltonian (\ref{ham_sec_quant}) and $E$ would be exact in the exact AC formulation.

The AC formula linearly interpolates between the zeroth-order Hamiltonian $\hat{H}^{(0)}$ and the exact one, $\hat{H}$ (\ref{ham_sec_quant})

\begin{eqnarray}
  \hat{H}^{\alpha} = \hat{H}^{(0)} + \alpha \hat{H}^{\prime}, \quad \text{with} \quad \hat{H}^{\prime} = \hat{H} - \hat{H}^{(0)} \label{h_alpha} \\ \text{and} \quad \alpha:0 \rightarrow 1 \nonumber
\end{eqnarray}

\noindent
where $\hat{H}^{(0)}$ can be either the group product function Hamiltonian~\cite{Pernal2018,Pastorczak2018} or also the Dyall Hamiltonian~\cite{Matousek2023}.

By exploiting the Hellmann-Feynman theorem, the exact AC correlation energy formula reads as

\begin{eqnarray}
  E^{\text{AC}}_{\text{corr}} & = & \int_0^1 \tilde{W}^{\alpha} \rm{d}\alpha \label{ac_corr} \\
  \tilde{W}^{\alpha} & = & \bra{\Psi^{\alpha}} \hat{H}^{\prime} \ket{\Psi^{\alpha}} \nonumber \\
  & - & \bra{\Psi_{\text{CAS}}} \hat{H}^{\prime} \ket{\Psi_{\text{CAS}}}
  \label{W}
\end{eqnarray}

\noindent
where $\tilde{W}^{\alpha}$ is the exact AC integrand and $\Psi^{\alpha}$ denotes the ground state of $H^{\alpha}$ (\ref{h_alpha}). The exact formulation is certainly impractical and a series of approximations have to be employed in order to transform (\ref{ac_corr}) into the practical form~\cite{Pernal2018, Pastorczak2018}. In the first place, all approximate AC methods assume that 1-RDM, $\gamma$, stays constant along the AC path (so called fixed-RDM approximation~\cite{Matousek2023}), because AC corrections account for (mainly) dynamical electron correlation, which may alter 1-RDM only marginally. Moreover, the extended random phase approximation (ERPA)~\cite{Chatterjee2012}, have been used to approximate the $\alpha$-dependent one-electron transition density matrices, $\gamma^{\alpha, 0 \nu}_{pq} = \bra{\Psi^{\alpha}_0} \hat{a}^{\dagger}_p \hat{a}_q \ket{\Psi^{\alpha}_{\nu}}$, which appear in (\ref{W}) after applying the exact relation between the 2-RDMs and one-body reduced functions: 1-RDMs and transition-1-RDMS~\cite{Pernal2018}. 

The general ERPA equations~\cite{Chatterjee2012}
are derived from the Rowe's equations of motion (EOM)~\cite{Rowe1968} under the assumption that a given excited state $\Psi^{\alpha}_{\nu}$ is obtained from the ground state $\Psi^{\alpha}_0$ by action of an excitation operator which includes only single excitations

\begin{eqnarray}
  & \hat{O}^{\dagger}_{\nu} \ket{\Psi^{\alpha}_0} = \ket{\Psi^{\alpha}_{\nu}} \\
  & \hat{O}^{\dagger}_{\nu} = \sum_{p<q} \Big( X^{\alpha}_{pq} \hat{a}^{\dagger}_p \hat{a}_q + Y^{\alpha}_{pq} \hat{a}^{\dagger}_q \hat{a}_p \Big) 
  \label{O_single}
\end{eqnarray}

\noindent
We would like to note that a related EOM approach, so called quantum EOM (qEOM)~\cite{Ollitrault2020, Asthana2023,pavosevic2023spinflip}, have been used for calculations of excitation energies on a quantum computer.

The indices $p$ and $q$ in (\ref{O_single}) naturally go over the full orbital space (inactive, active, and virtual) and solution of the $\alpha$-dependent ERPA equations would thus require $\alpha$-dependent 1- and 2-RDMs, which are approximated by the reference ($\alpha = 0$) RDMs within the fixed-RDM approximation. Moreover, due to the structure of the reference CAS wave function, the RDMs with general indices can be constructed solely from active space RDMs (with all active indices). Consequently, the AC approximation requires only active space 1- and 2-RDMs, quantities which are directly accessible in the VQE procedure.

It was demonstrated numerically in Refs.~\citenum{Pastorczak2018, Pastorczak2019} that one can avoid the integration in (\ref{ac_corr}) by linearized-AC-integrand approximation, named AC0, without losing much accuracy. Instead of solving the full ERPA problem as in AC, only
the smaller-sized ERPA equations for specific blocks (active-active, active-inactive, virtual-active, virtual-inactive) at $\alpha = 0$ have to be solved in the case of AC0. This results in an overall scaling $n^2_{\text{virt}} n^4_{\text{act}}$, $n_{\text{virt}} n^5_{\text{act}}$, $n^6_{\text{act}}$, where $n_{\text{act}}$ denotes the number of active orbitals and $n_{\text{virt}}$ denotes the number of virtual orbitals. The computational complexity of the full AC problem can be reduced resulting in the scaling with the 5th power of the system size, at fixed number of the active orbitals, with the help of Cholesky decomposition~\cite{Drwal2022}. Recently, we have combined AC methodology with the density matrix renormalization group (DMRG) algorithm~\cite{Beran2021}, in which DMRG is responsible for a proper description of the strong correlation, whereas dynamical correlation is computed via the AC technique.

\subsection{Orbital Optimization}

As mentioned in the previous subsection, the AC theory relies in its derivation on the Hellmann-Feynman theorem, i.e. it expects the reference wave function to be optimal with respect to all variational parameters, in our case UCC amplitudes as well as molecular orbital (MO) coefficients. The standard UCCSD-VQE method (or its adaptive improvements) do not fulfill this requirement. Nevertheless, orbital-optimized (OO) UCC-VQE~\cite{Takeshita2020, Sokolov2020, Mizukami2020, Yalouz2021, deGraciaTrivio2023},  as well as ADAPT-VQE-SCF~\cite{adapt_vqe_scf} has been developed recently.

The orbital rotation (unitary change of MO basis) can be expressed as an exponential of an antihermitian operator, $\hat{\kappa}$~\cite{olsen_bible}

\begin{eqnarray}
  \hat{U}(\vec{\kappa}) & = & e^{\hat{\kappa}} \label{exp_kappa} \\
  \hat{\kappa} & = & \sum_{pq} \kappa_{pq} (\hat{E}_{pq} - \hat{E}_{qp}) \\
   & = & \sum_{pq} \kappa_{pq} \hat{E}_{pq}^{-}
\end{eqnarray}

\noindent
where $\vec{\kappa} = \{\kappa_{pq}\}$ is a vector of orbital parameters and $\hat{E}_{pq}$ is a singlet excitation operator

\begin{equation}
  \hat{E}_{pq} = \hat{a}^{\dagger}_{p\alpha} \hat{a}_{q \alpha} + \hat{a}^{\dagger}_{p\beta} \hat{a}_{q \beta}
\end{equation}

\noindent
and $p,q,\ldots$ denote general MO indices.
The orbital optimization is equivalent to the energy minimization with respect to orbital rotation parameters, $\vec{\kappa}$. 

The OO-UCC energy can be expressed as

\begin{small}
\begin{eqnarray}
  E(\vec{\theta}, \vec{\kappa}) & = & \bra{\Psi(\vec{\theta})} e^{-\hat{\kappa}} \hat{H} e^{\hat{\kappa}} \ket{\Psi(\vec{\theta})} \nonumber \\
  & = & \bra{\Psi_{\text{ref}}} \hat{U}(\vec{\theta})^{\dagger} e^{-\hat{\kappa}} \hat{H} e^{\hat{\kappa}} \hat{U}(\vec{\theta}) \ket{\Psi_{\text{ref}}}
\end{eqnarray}
\end{small}

\noindent
and its Taylor expansion to the second order in the variational parameters would lead to the second-order Newton-Raphson procedure, which couples optimization of UCC amplitudes and MO coefficients via the mixed second derivatives $\partial^2 E/ \partial \theta_i \partial \kappa_j$~\cite{Sun2017, adapt_vqe_scf}.

In this work, we use a simpler and computationally cheaper two-step approach, which at the cost of slower convergence neglects the mixed derivatives and proceeds in the decoupled inner-outer loop fashion. In the inner loop, the UCC amplitudes are variationally optimized and then, in the outer loop, for the optimized amplitudes, the gradient, $g_{pq}$ (and possibly also the Hessian, $H_{pq,rs}$) with respect to the orbital rotations can be computed

\begin{eqnarray}
  g_{pq} & = & \bra{\Psi(\vec{\theta})} [\hat{H}, \hat{E}^-_{pq}] \ket{\Psi(\vec{\theta})} \\
  H_{pq,rs} & = & \frac{1}{2} \bra{\Psi(\vec{\theta})} [[\hat{H}, \hat{E}^-_{pq}],\hat{E}^-_{rs}] + \nonumber \\
  & + & [[\hat{H}, \hat{E}^-_{rs}],\hat{E}^-_{pq}] \ket{\Psi(\vec{\theta})}
\end{eqnarray}

\noindent
We have employed the quasi-Newton Broyden–Fletcher–Goldfarb–Shanno (BFGS) procedure with gradient preconditioning as described in Ref.~\citenum{Levine2020}. Once the new set of MOs is obtained via the BFGS method, the inner loop continues with the new UCC amplitude optimization and the whole procedure is repeated until convergence.

Since the structure of the UCC single excitations is identical to the orbital rotation operator (\ref{exp_kappa}), it can be absorbed into the classical orbital-optimization routine, which was briefly described above, and only the double excitations remain in the UCC cluster operator~(\ref{ucc_anz}) \cite{Mizukami2020}. This method is denoted as OO-UCCD.

The combination of the orbital optimized VQE algorithms and the AC methodology is rather straightforward, however, important issues arise, which are related to the fact that  
neither UCCD-VQE, nor ADAPT-VQE (with the limited number of iterations) correspond exactly to FCI.
Firstly, the orbital rotations comprising only the active indices are not redundant and should be, in principle, included in the optimization~\cite{olsen_bible}. 
Secondly, wave functions provided by OO-UCCD or ADAPT-VQE-SCF are not the exact eigenfunctions of the zero-order Hamiltonians (\ref{h_alpha}) and the question is how well AC performs in these situations.
Moreover, since AC relies on the orbital-optimized wave functions, the question remains as to how the accuracy of orbital optimization when employing the adaptive methods will affect the resulting AC energies. In the present study, we have addressed the aforementioned issues by means of classical numerical simulations of challenging electronic structure problems presented below.

\section{Computational Details}

\begin{figure}[!ht]
  \includegraphics[width=6cm]{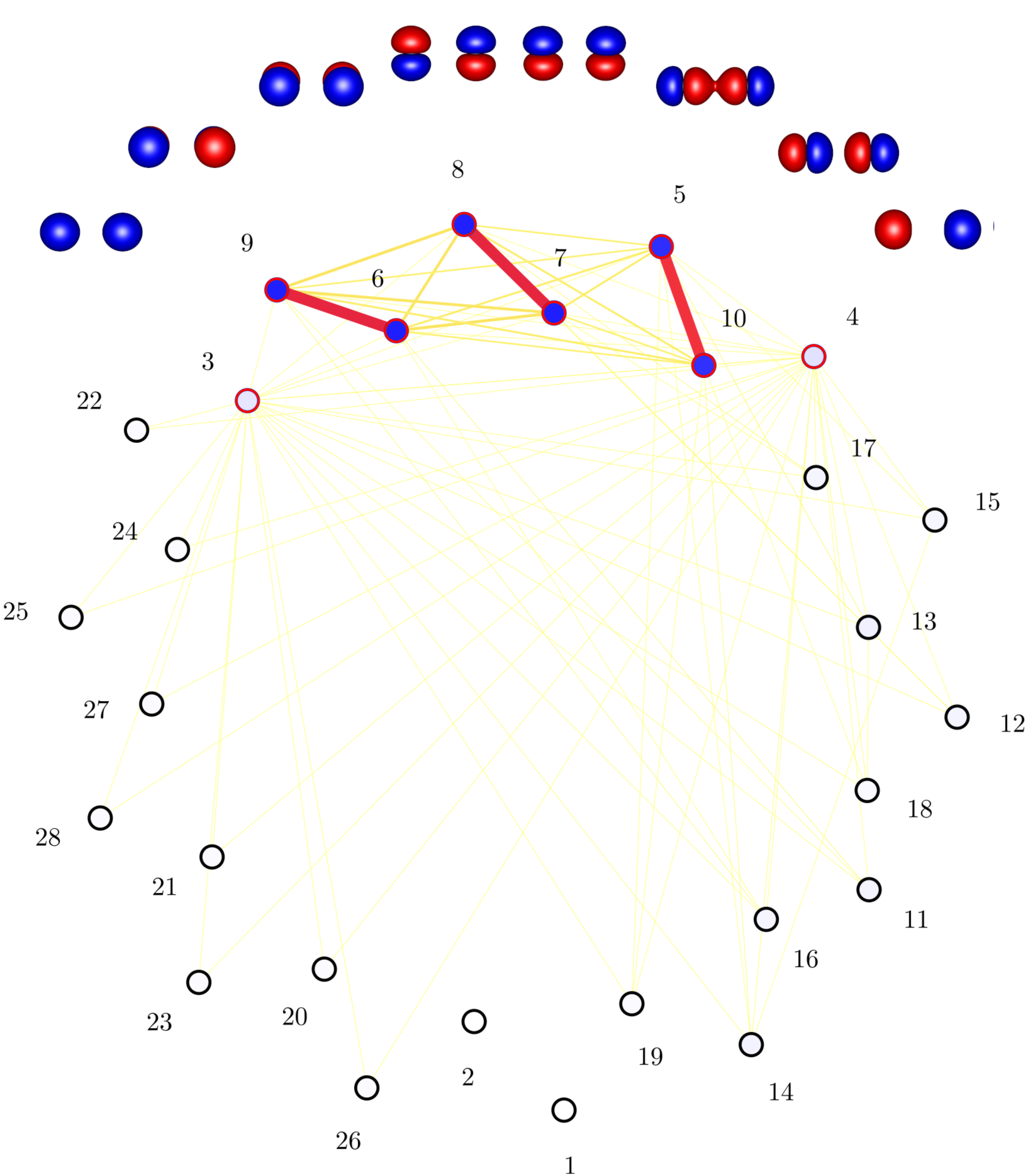}
  \includegraphics[width=6cm]{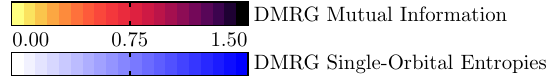}
  \caption{DMRG mutual information (colored edges) and single-orbital entropies (colored vertices) of N$_2$ molecule for  $r = 2.5$ \AA . Red circles represent the orbitals with $s_i > 0.1$, which are also depicted.}
  \label{fig_n2_mutual}
\end{figure}

\begin{figure*}[!ht]
  \subfloat[\label{tme_reaction}]{%
    \includegraphics[width=8cm]{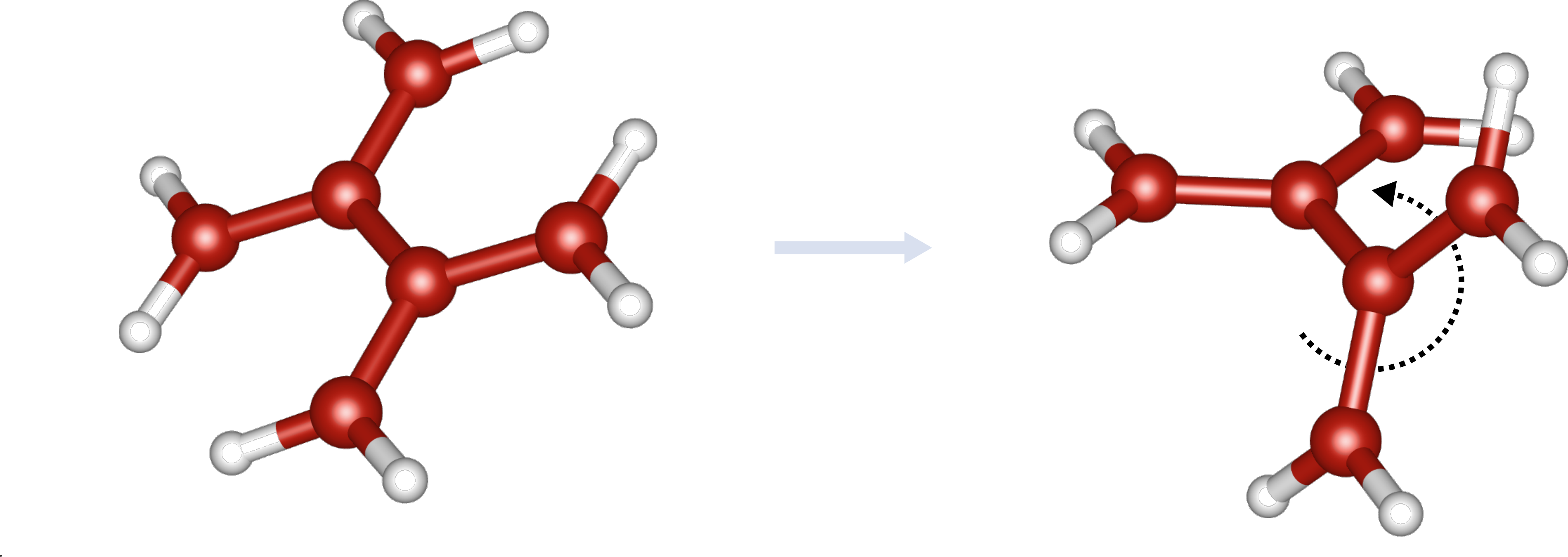}
  } \\
  \subfloat[\label{tme}]{%
    \includegraphics[width=9cm]{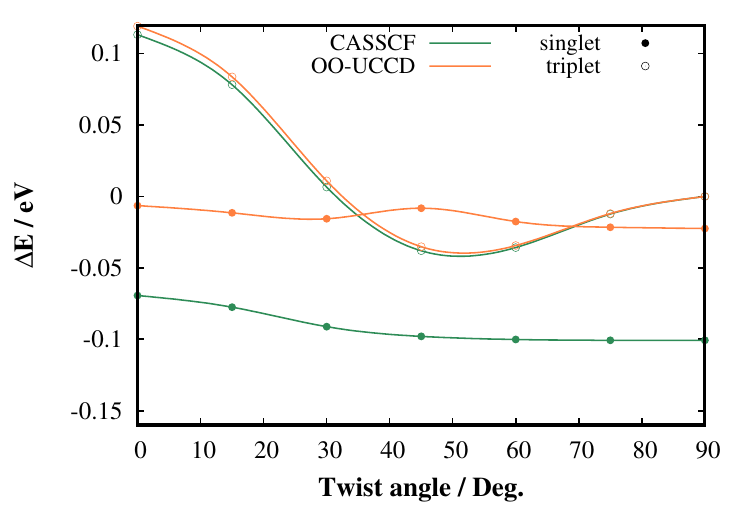}
  }
  \subfloat[\label{tme_ac0}]{%
    \includegraphics[width=9cm]{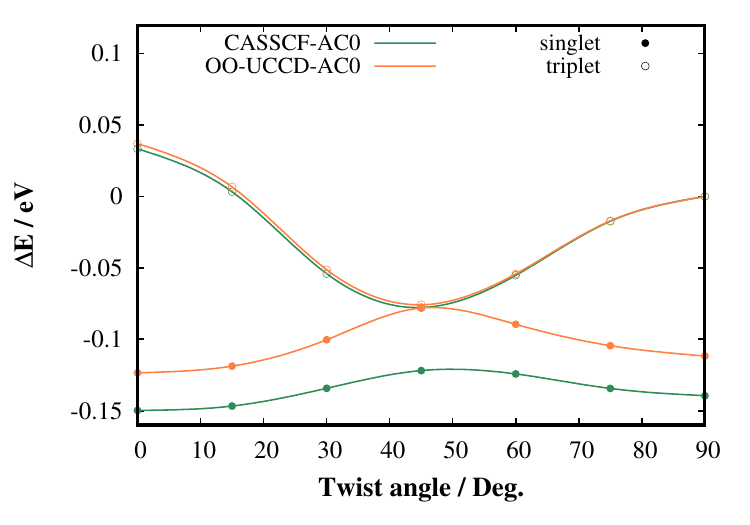}
  }
  \caption{(a) Studied process of a rotation of the TME allyl subunits about the central C-C bond. Carbon atoms are colored red; hydrogens are white. (b) CASSCF(6,6) and OO-UCCD(6,6), and (c) CASSCF(6,6)-AC0 and OO-UCCD(6,6)-AC0 singlet and triplet state twisting PESs in cc-pVTZ basis.}
  \label{fig_tme}
\end{figure*}

As mentioned in the introduction, we studied two strongly correlated problems, namely the dissociation of N$_2$, which corresponds to breaking of the triple bond, and the twisting of the TME biradical about the central C-C bond (Figure \ref{tme_reaction}). 
The latter is of utmost importance for this study, since even just a qualitatively correct description of the TME twisting process requires an adequate treatment of both the strong (static), as well as dynamical electron correlation~\cite{Pozun2013}.

\begin{figure*}[!ht]
  \subfloat[\label{fig_n2_dz_8orbs}]{%
    \includegraphics[width=9cm]{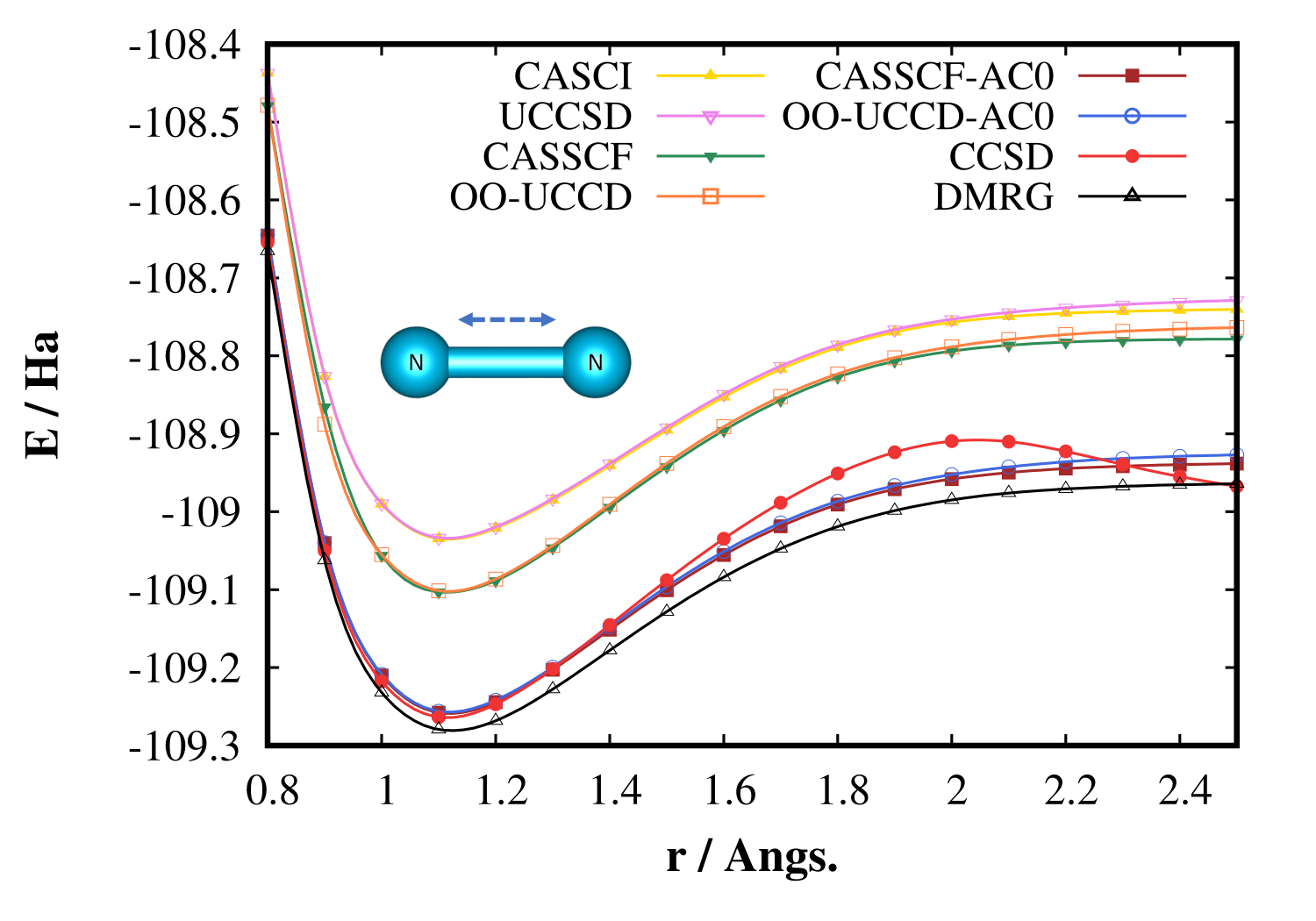}
  }
  \subfloat[\label{fig_n2_dz_corr_full}]{%
    \includegraphics[width=9cm]{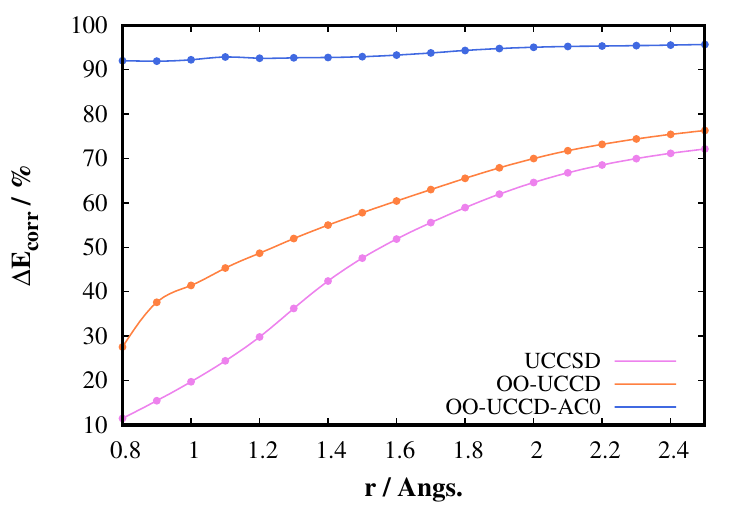}
  }
  \caption{(a) Dissociation energy curves of N$_2$ molecule computed by different methods in cc-pVDZ basis. The CASCI, UCCSD, CASSCF, and OO-UCCD methods were restricted to CAS(10,8). (b) Percentage of the correlation energy ($E_{\text{DMRG}}$ - $E_{\text{HF}}$) retrieved.}
  \label{fig_n2}
\end{figure*}

\begin{figure*}[!ht]
  \subfloat[\label{fig_n2_dz_err_8orbs}]{%
    \includegraphics[width=9cm]{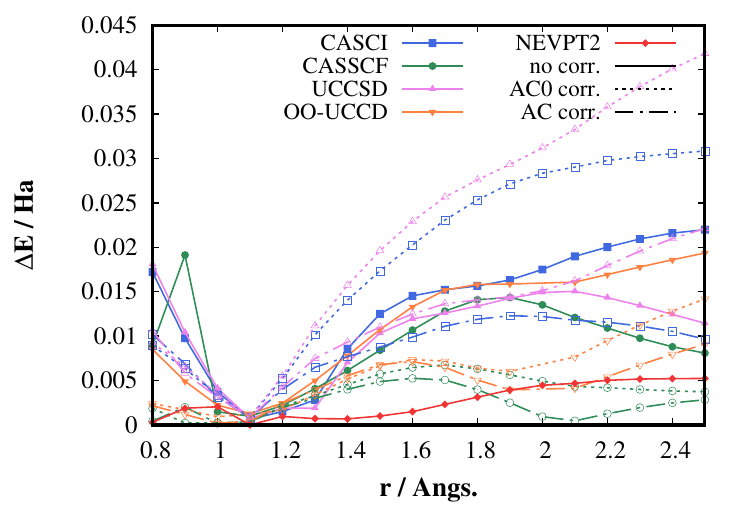}
  }
  \subfloat[\label{fig_n2_dz_act}]{%
    \includegraphics[width=9cm]{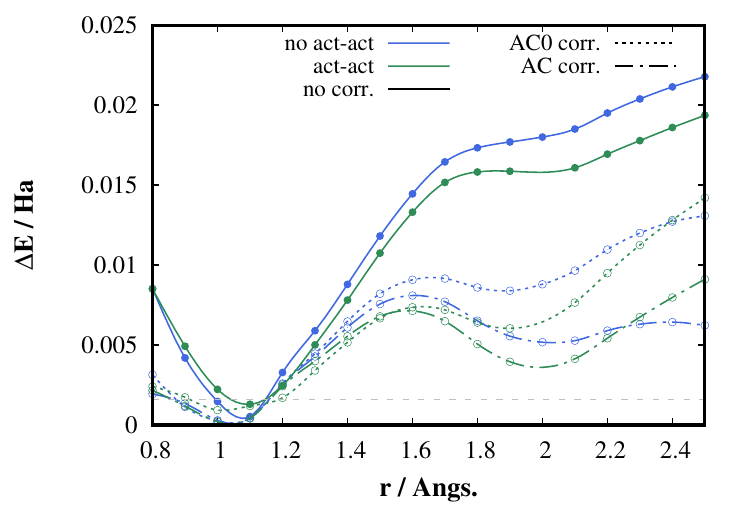}
  }  
  \caption{(a) Absolute errors of relative energies ($E_{\text{rel}}=E - E_{\text{min}}$) with respect to the exact (DMRG) results of N$_2$ molecule computed by different methods in cc-pVDZ basis. (b) The effect of active-active rotations in OO-UCCD orbital optimization on AC0/AC energies. The dashed horizontal grey line corresponds to the chemical accuracy, i.e. the error of 1 kcal/mol.}
  \label{fig_n2_rel}
\end{figure*}

In the case of N$_2$ molecule, we used the cc-pVDZ basis~\cite{Dunning1989}, in which the reference extrapolated DMRG energies of FCI quality are feasible. 
We extended the standard active space comprising six electrons in two $\sigma$ and four $\pi$ orbitals, CAS(6,6), to CAS(10,8) by including two doubly occupied $\sigma$ MOs composed of 2s orbitals. The contribution of the latter to correlation is for longer internuclear distances not negligible. As it can be seen in Fig. \ref{fig_n2_mutual}, for $r=2.5$~\AA, their single-orbital entropies, which quantify the importance for the active space~\cite{Stein2016}, are larger than 0.1. It turns out that the fixed-RDM approximation is more justifiable in this larger CAS(10,8) and 
the AC corrections are more accurate. This is in agreement with Ref.~\citenum{Pastorczak2018}, in which we showed on the same N$_2$ example the systematic improvement of the CASSCF-AC0/AC results with enlarging active spaces.

The TME calculations were performed in the cc-pVTZ basis~\cite{Dunning1989}, which was shown by Pozun et al.\ in Ref.~\citenum{Pozun2013} to be the minimal basis providing enough flexibility for proper description of the twisting process. The geometries of the TME biradical along the twisting process were taken from Ref.~\citenum{Veis2016}. The CAS comprised all $\pi$ orbitals, i.e.\ 6 electrons in 6 orbitals, CAS(6,6).

The UCC calculations were performed with our in-house C++ circuit-based quantum computer simulator which was used previously e.g. for simulated QPE computations~\cite{Veis2010, Veis2012, Veis2016}, adiabatic state preparation~\cite{Veis2014}, and most recently also for VQE simulations~\cite{Bauman2021}. The quantum computer simulator was interfaced to the local version of \textsf{Orca}~\cite{orca}, whose CASSCF routines were extended to include active-active rotations, for the purposes of OO-UCC simulations. The AC0/AC corrections were computed with the \textsf{GammCor} program~\cite{gammcor}. All supporting quantum chemical calculations were performed in \textsf{Orca}~\cite{orca}, except of the DMRG ones, which were carried out in \textsf{MOLMPS}~\cite{Brabec2020}.

The ADAPT-VQE-SCF as well as qubit-ADAPT-VQE-SCF simulations were performed with the \textsf{Qiskit}~\cite{Qiskit} \texttt{AdaptVQE} routine, which was similarly to OO-UCC interfaced to \textsf{Orca}. We employed the parity fermion-to-qubit mapping~\cite{Seeley2012} and a single Trotter step in all \textsf{Qiskit} simulations. We also used the tapering-off technique based on $Z_2$ symmetries \cite{Bravyi_2017, Setia2020}, which 
reduced the number of qubits in TME simulations from 12 to 8. In the present work, we have performed the noise-free state-vector simulations. The effects of noise will be the subject of the follow-up study.
 
\section{Results}

\subsection{Nitrogen molecule}

First, we present the results for the nitrogen molecule, which are collected in Figs.~\ref{fig_n2} and~\ref{fig_n2_rel}. Dissociation of N$_2$ is a notoriously known strongly correlated problem, which single-reference methods such as CCSD, see Fig.~\ref{fig_n2}, or CCSD(T) fail to describe. The UCCSD method, unlike CCSD, provides a qualitatively correct shape of the potential energy surface (PES), however, when restricted only to the active space, in our case CAS(10,8), the energy is too high due to missing dynamical correlation energy. 

As can be seen in Fig.~\ref{fig_n2_dz_8orbs}, the UCCSD PES parallels closely the CASCI one for shorter internuclear distances and starts to deviate for longer distances, where the multireference character of the ground state wave function is more pronounced. The orbital optimization brings some amount of the out-of-CAS electron correlation and decreases the energy considerably. Similarly to UCCSD, the OO-UCCD PES follows the CASSCF PES for shorter distances and deviates for longer ones. 
Most importantly, the AC0 correction on top of OO-UCCD provides energies much closer to the reference DMRG values. Moreover, the difference between the OO-UCCD-AC0 and CASSCF-AC0 energies is for longer distances lower than the difference between the non-corrected energies (11 vs 15 mHa for 2.5 \AA), which numerically confirms that OO-UCCD can be successfully used to provide the zero-order wave functions for the AC methodology.
Figure~\ref{fig_n2_dz_corr_full} shows the amount of the correlation energy, defined as the difference between the exact, i.e.\ DMRG energy, and the HF energy, retrieved by the individual methods. One can see, that OO-UCCD-AC0 recovers more than 90 $\%$ of the correlation energy for the whole range of internuclear distances and dramatically improves the UCCSD and OO-UCCD results.

The absolute values of errors of the relative energies, $E_{\text{rel}} = E - E_{\text{min}}$, with respect to the DMRG relative energies are depicted in Fig.~\ref{fig_n2_dz_err_8orbs}. In this detailed plot, not only AC0, but also the full AC results are presented. The Figure well demonstrates the importance of orbital optimization, which is a presumption of the AC methodology. The AC0 corrected results of non-orbital optimized methods (CASCI, UCCSD) are in fact considerably worse than the original non-corrected results. The full AC correction works better in this respect and improves upon CASCI, but UCCSD-AC is worse than UCCSD for $r>2$~\AA. As expected, the more sophisticated as well as expensive AC method provides more accurate results than its linearized integrand approximation AC0, nevertheless both corrections systematically and considerably improve upon OO-UCCD (as well as CASSCF). The AC results are only slightly worse than the results of the strongly contracted NEVPT2 method, which works accidentally very well in this case \cite{Pastorczak2018}. One should, however, keep in mind that NEVPT2 is more costly than AC0/AC, since it requires up to 4-electron active space RDMs~\cite{Tammaro2023}.

In Fig.~\ref{fig_n2_dz_act}, we show the effect of active-active rotations on the performance of \mbox{OO-UCCD(-AC0/AC)} methods. Their importance is not surprisingly increasing with the increasing internuclear distance, since UCCD deviates more from FCI due to the stronger multireference character. The active-active rotations decrease the OO-UCCD energy for $r=2.5$~\AA~by about 2.4 mHa, thus not negligibly. The AC0/AC corrected energies are affected similarly, but the trend is not systematic. The role of active-active rotations is expected to be more pronounced for larger and more correlated active spaces.
In case of the TME biradical discussed below, the effect of active-active rotations was smaller (at the order of $\mu$Ha) due to the smaller CAS(6,6).

\subsection{Tetramethyleneethane}

\begin{figure*}[!ht]
  \subfloat[\label{adapt_ferm}]{%
    \includegraphics[width=6cm]{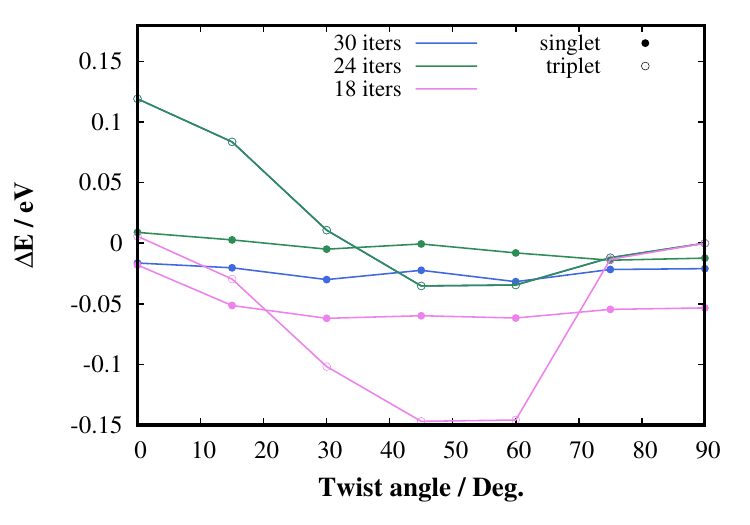}
  }
  \subfloat[\label{adapt_ferm_ac0}]{%
    \includegraphics[width=6cm]{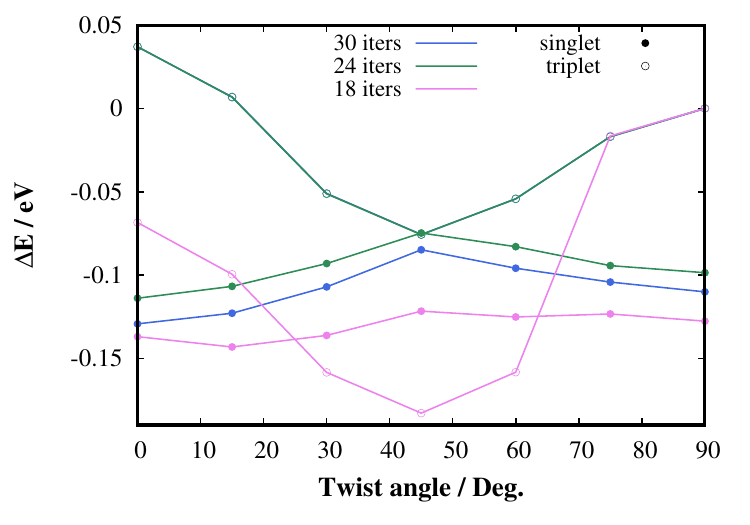}
  }
  \subfloat[\label{adapt_cnots_ferm}]{%
    \includegraphics[width=6cm]{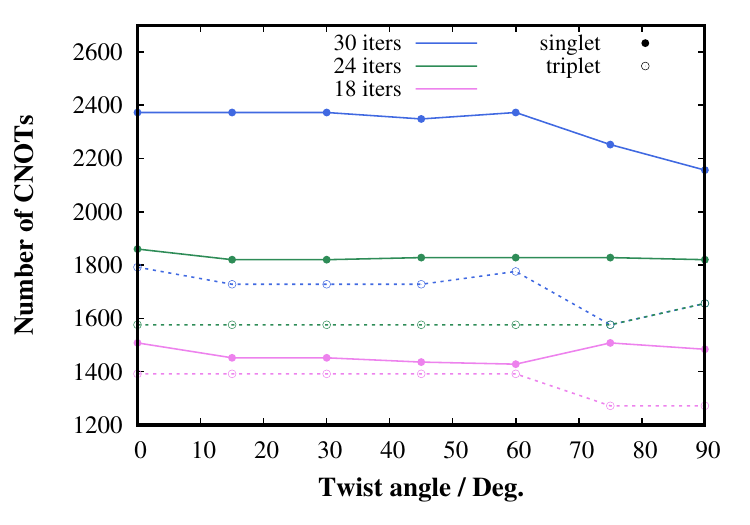}
  }
  \caption{Singlet and triplet state twisting PESs of TME molecule in cc-pVTZ basis calculated by (a) ADAPT-VQE-SCF, (b) ADAPT-VQE-SCF-AC0. (c) ADAPT-VQE CNOT gate counts. The results corresponding to 18, 24, and 30 iterations of the adaptive procedure are displayed.}
  \label{fig_adapt}
\end{figure*}

\begin{figure*}[!ht]
  \subfloat[\label{adapt}]{%
    \includegraphics[width=6cm]{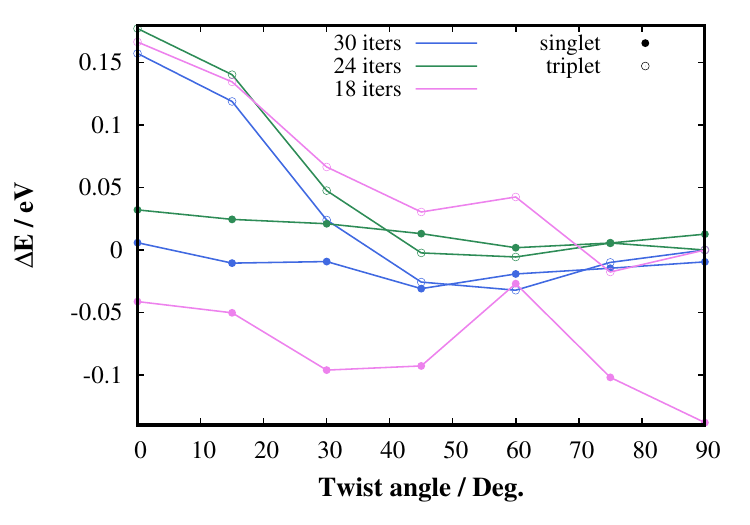}
  }
  \subfloat[\label{adapt_ac0}]{%
    \includegraphics[width=6cm]{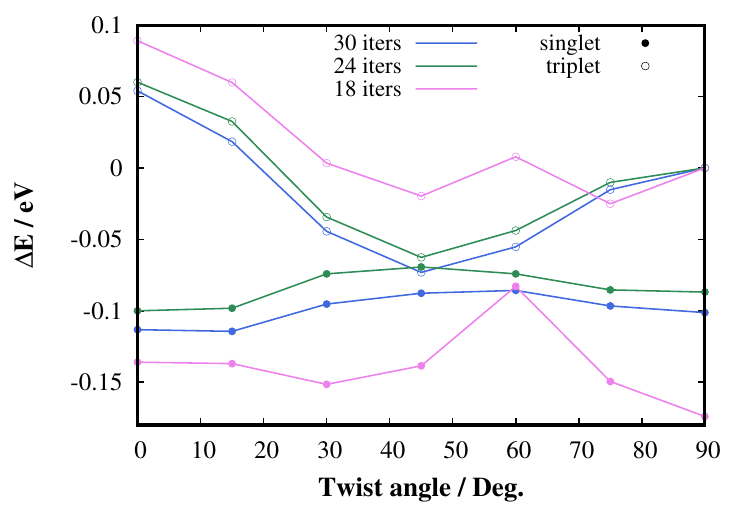}
  }
  \subfloat[\label{adapt_cnots}]{%
    \includegraphics[width=6cm]{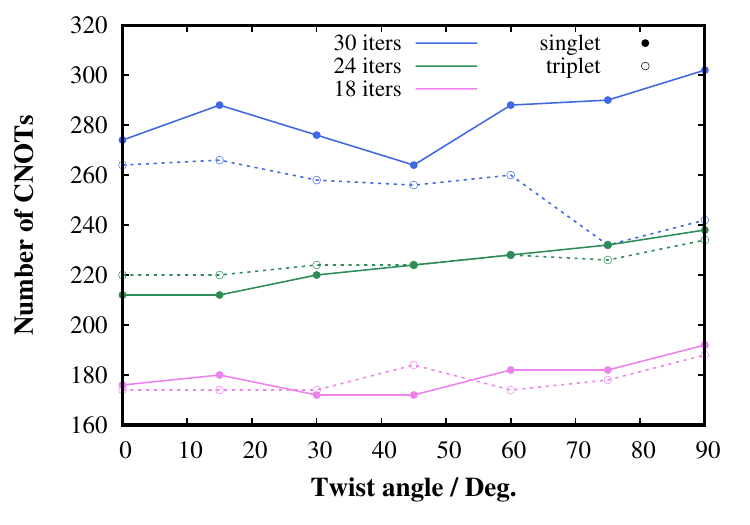}
  }
  \caption{Singlet and triplet state twisting PESs of TME molecule in cc-pVTZ basis calculated by (a) qubit-ADAPT-VQE-SCF, (b) qubit-ADAPT-VQE-SCF-AC0. (c) qubit-ADAPT-VQE CNOT gate counts. The results corresponding to 18, 24, and 30 iterations of the adaptive procedure are displayed.}
  \label{fig_qubit_adapt}
\end{figure*}

The second problem on which we have tested the combined VQE-AC methodology is the twisting process of the TME biradical depicted in Figure~\ref{tme_reaction}. TME is the simplest disjoint non-Kekul\'{e} biradical and a very intricate system, indeed. It was shown by Pozun et al.~\cite{Pozun2013}, that the correct shape and ordering of the singlet and triplet PES corresponding to a free rotation of the TME allyl subunits about the central single C-C bond requires: (a) flexible-enough atomic basis set, at least of a triple-$\zeta$ quality, (b) all-$\pi$ active space, i.e. CAS(6,6), for a correct description of the strong correlation, (c) dynamical (out-of-CAS) electron correlation described at least at the PT2 level (or similar). The last mentioned is needed to get the singlet PES with an energy maximum at 45$^{\circ}$, where the triplet state, which is higher in energy, has its minimum. The energy gap between the two states at 45$^{\circ}$ is only about 0.02~eV~\cite{Pozun2013, tme_paper}. Consequently, determining the relative stability of both states turned out to be a challenge for experimental as well as theoretical methods. In our view, the TME biradical is due to its properties a perfect candidate for benchmarking of new quantum algorithms for NISQ devices.

\begin{figure*}[!ht]
  \subfloat[\label{adapt_conv}]{%
    \includegraphics[width=9cm]{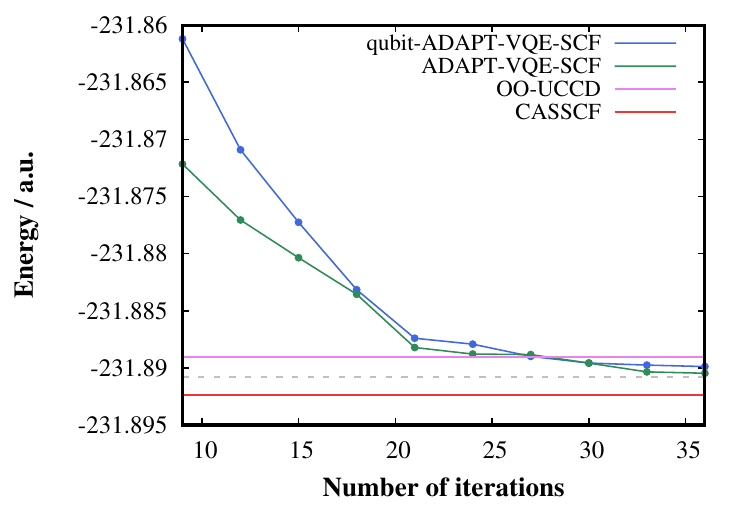}
  }
  \subfloat[\label{adapt_triplet_conv}]{%
    \includegraphics[width=9cm]{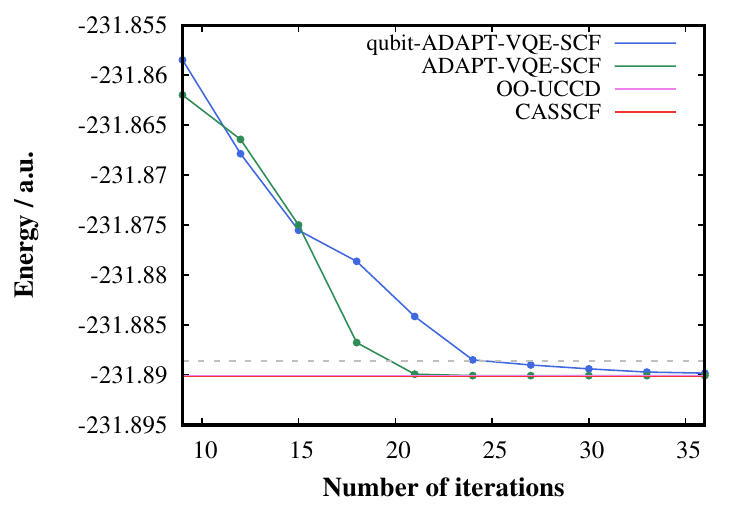}
  } 
  \caption{Convergence of ADAPT-VQE-SCF and qubit-ADAPT-VQE-SCF with respect to the number of iterations in the adaptive procedure for the (a) singlet state and (b) triplet state of TME in the geometry with the twist angle of 45$^{\circ}$. The dashed horizontal grey lines correspond to the chemical accuracy, i.e. the error of 1 kcal/mol with respect to CASSCF.}
  \label{fig_adapt_conv}
\end{figure*}

\begin{figure}[!ht]
  \includegraphics[width=9cm]{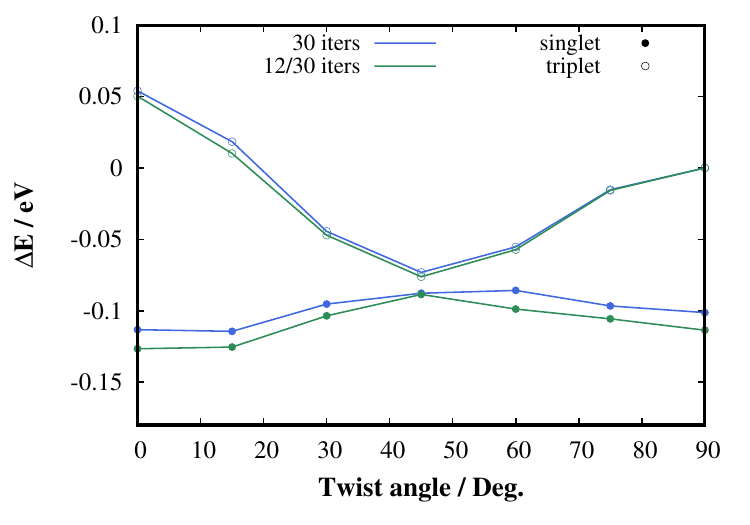}
  \caption{Comparison of qubit-ADAPT-VQE-SCF-AC0 singlet and triplet state twisting PESs of the TME molecule employing 30 iterations of the adaptive procedure with the same method employing only 12 iterations for the orbital optimization part and 30 iterations for RDMs generation for the AC0 correction.}
  \label{fig_adapt_test}
\end{figure}

We used the cc-pVTZ basis and we restricted ourselves to only the AC0 method. Moreover, motivated by a possible future physical realization on NISQ hardware, we tested also the adaptive VQE approaches, which implementation requires fewer CNOT gates than the UCCSD. The UCC results are presented in Fig.~\ref{fig_tme}, whereas the results of the adaptive VQE approaches are collected in Figures \ref{fig_adapt} - \ref{fig_adapt_test}. The comparison of the key features of the TME twisting process computed by different methods with the best available experimental and theoretical results is shown in Table~\ref{tab_tme}.

Let us first discuss the UCC results. 
In case of TME, we present only the orbital optimized results since the orbital optimization improves the shapes of both PESs considerably. As can be seen in Fig.~\ref{tme}, CASSCF(6,6)/cc-pVTZ provides correct ordering of both states (singlet is energetically lower), correct shape of the triplet PES, however, the singlet PES is lacking the characteristic ``bump'' (maximum at 45$^{\circ}$), which was attributed to dynamical electron correlation effects by Pozun et al.~\cite{Pozun2013}. The agreement between the triplet OO-UCCD and CASSCF PESs is excellent, which is a consequence of the single reference nature of this state. On the contrary, both methods compare a little worse for the singlet state, which is strongly correlated. Most importantly, singlet is incorrectly higher in energy than triplet around 45$^{\circ}$ at the OO-UCCD level. Moreover, somewhat interestingly the singlet PES evinces a small ``bump'', which probably stems from cancellation of errors, because dynamical (out-of-CAS) correlation effects are not included.

Figure~\ref{tme_ac0} compares the AC0 corrected CASSCF(6,6) and OO-UCCD(6,6) PESs. One can see that the dynamical electron correlation causes a development of the ``bump'' on the singlet PES at 45$^{\circ}$. Again, both methods coincide for the triplet state, but the OO-UCCD-AC0 curve is noticeably higher in energy than the CASSCF-AC0 one for the singlet state. Nevertheless similarly to the N$_2$ example, the difference between OO-UCCD-AC0 and CASSCF-AC0 is much smaller than the difference between OO-UCCD and CASSCF (0.04~eV vs 0.09~eV for 45$^{\circ}$). When comparing the final OO-UCCD-AC0 results with the best available experimental or theoretical results shown in Table~\ref{tab_tme}, one can observe an overall excellent agreement with the errors below  0.02~eV ($\approx 0.4$~kcal/mol), safely within the chemical accuracy.

\begin{table*}[!ht]
\begin{small}
\begin{tabular}{c c c c}
 & $\Delta E_{\text{T-S}}$/45$^{\circ}$ & $\Delta E_{\text{T-S}}$/90$^{\circ}$ & $\Delta E_{\text{twist}}$ \\
 & eV & eV & kcal/mol \\
\hline 
CASSCF & 0.06 & 0.10 & -0.660 \\
OO-UCCD & -0.03 & 0.02 & -0.041 \\
ADAPT-VQE-SCF & -0.01 & 0.02 & -0.139 \\
Qubit-ADAPT-VQE-SCF & 0.005 & 0.009 & -0.846 \\
CASSCF-AC0 & 0.04 & 0.14 & 0.643 \\
OO-UCCD-AC0 & 0.002 & 0.11 & 1.047 \\
ADAPT-VQE-SCF-AC0 & 0.009 & 0.11 & 1.026 \\
Qubit-ADAPT-VQE-SCF-AC0 & 0.015 & 0.10 & 0.590 \\ 
\hline
best estimate & 0.02\footnote{DMC result \cite{Pozun2013}} & 0.13 $\pm$ 0.013\footnote{Photoelectron spectroscopy result \cite{Clifford1998}} & 1.132 \footnote{FCIQMC results \cite{tme_paper}}
\end{tabular}
\caption{TME singlet-triplet energy gaps corresponding to the torsional angles of 45° and 90° and twisting energy barrier [E(45$^\circ$) - E(0$^\circ$)] in the singlet state calculated by different methods. The results of the adaptive VQE approaches with 30 iterations are displayed. \label{tab_tme}}
\end{small}
\end{table*}

We now compare the performance of the AC0 corrected adaptive VQE approaches with the view of a possible future physical implementation on quantum devices. In Fig.~\ref{fig_adapt}, the results of the ADAPT-VQE-SCF method~\cite{Grimsley2019, adapt_vqe_scf} with the underlying VQE subroutine employing the pool of fermionic excitation operators are presented, whereas in Fig.~\ref{fig_qubit_adapt} the results of the qubit-ADAPT-VQE-SCF~\cite{Tang2021, adapt_vqe_scf} method with the pool of Pauli operators are shown. We display the representative examples with 18, 24, and 30 iterations of the adaptive build of the VQE ansatz, which cover the transition from ``not accurate'' to ``accurate enough'' regime. In Figs.~\ref{adapt_ferm} and~\ref{adapt_ferm_ac0}, one can see that 24 iterations of ADAPT-VQE, which corresponds to about 1800 CNOT gates for the singlet state (see Fig.~\ref{adapt_cnots_ferm}), is sufficient to get the ADAPT-VQE-SCF PESs of the same quality as the OO-UCCD ones and consequently ADAPT-VQE-SCF-AC0 equivalent to OO-UCCD-AC0 (compare with Fig.~\ref{tme}). Notice that the triplet state curves corresponding to 24 and 30 iterations overlap. Moreover, 30 iterations of ADAPT-VQE provide the singlet state results of slightly better quality and the resulting ADAPT-VQE-SCF-AC0 features of the singlet and triplet state PESs in excellent agreement with the benchmark numbers, as can be seen in Table~\ref{tab_tme}.

The qubit-ADAPT-VQE method was proposed as a CNOT gates saving alternative to ADAPT-VQE~\cite{Tang2021}. One can see in Fig.~\ref{adapt} that the qubit adaptive procedure with 30 iterations provides results of sufficient quality with an order of magnitude less CNOT gates. Fewer than 300 CNOTs suffices to achieve the qubit-ADAPT-VQE-SCF-AC0 relative energy curves whose main features safely fall within the chemical accuracy difference with respect to the benchmark results (see Table~\ref{tab_tme}). The only visible weak point is a less pronounced ``bump'' on the singlet PES.

In Fig.~\ref{fig_adapt_conv}, we demonstrate the convergence of \mbox{(qubit-)ADAPT-VQE-SCF} energy with the number of iterations compared to CASSCF and OO-UCCD energies for the geometry with the twist angle of $45^{\circ}$. One can see that in case of the strongly correlated singlet state, both \mbox{ADAPT-VQE-SCF} as well as \mbox{qubit-ADAPT-VQE-SCF} get closer to CASSCF than OO-UCCD, however, even for almost 40 iterations the error is still slightly higher than the chemical accuracy. Nevertheless, the final AC0 corrected relative energy gaps, which are measurable quantities, agree very well with the best available results, as discussed above.

We have also numerically tested whether the orbital optimization can be performed with a less accurate qubit-ADAPT-VQE CAS solver. As can be seen in Fig.~\ref{fig_adapt_test}, the qubit-ADAPT-VQE-SCF method with only 12 iterations of the adaptive build of the VQE ansatz (requiring about 100 CNOT gates), which is energetically very inaccurate (see Fig.~\ref{fig_adapt_conv}), is able to provide the optimized orbitals of a sufficient quality. Only a single qubit-ADAPT-VQE run with 30 iterations is then necessary
to obtain the accurate active space RDMs  for the subsequent AC0 correction.

\section{Conclusion}

In this work, we have addressed the electronic structure problem on near-term quantum computers. Due to the limited quantum resources, NISQ devices are aimed at the most difficult part of the problem, i.e. the manifold of strongly correlated orbitals, CAS.
Herein, we have presented the novel computational approach, which improves dramatically upon the VQE algorithms~\cite{Cao2019} by adding the dynamical electron correlation effects comprising electrons outside of CAS by means of the classical AC corrections~\cite{Pernal2018, Pastorczak2018}. 
Our approach does not bring any additional requirements on quantum resources as the AC corrections work with the active space 1- and 2-RDMs, quantities directly accessible in VQE. The classical computational demands of the AC corrections employed here are only modest and scale as $\mathcal{O}({n^6})$, which can be further reduced to $\mathcal{O}({n^5})$ with the Cholesky decomposition~\cite{Drwal2022}. We expect our approach to be more immune to noise than e.g. the NEVPT2 corrections~\cite{Tammaro2023}, which require measuring of up to 4-RDMs. 

We have tested the performance of the orbital optimized UCC-based VQE-AC0/AC methods by classical noiseless simulations of two challenging strongly correlated problems, namely the dissociation of N$_2$ and the twisting process of TME. Especially the latter one deserves attention, since the dynamical electron correlation effects are crucial for the qualitatively correct shape of the singlet PES. Our simulations of both the aforementioned problems are in excellent agreement with the benchmark data and numerically confirm that the OO-UCCD can be successfully used to provide the zero-order wave functions for AC.

Since we consider the TME biradical to be a perfect testing case for NISQ quantum algorithms, we have also performed the (qubit)-ADAPT-VQE-AC0 simulations, which revealed that about 200-300 CNOT gates would suffice for obtaining accurate PESs. In the follow-up work, we will focus on the effect of noise on the performance of the VQE-AC methods as well as extending of the developed method to larger molecular systems via quantum embedding techniques~\cite{beran2023projection,rossmannek2023quantum}.

\section*{Acknowledgment}
This work was supported by the Czech Science Foundation, the Charles University Grant Agency (Grant No. 218222), the Ministry of Education, Youth and Sports of the Czech Republic through the e-INFRA CZ (ID:90254), and the National Science Center of Poland, under Grant No. 2021/43/I/ST4/02250.

\bibliography{references.bib}

\providecommand{\latin}[1]{#1}
\makeatletter
\providecommand{\doi}
  {\begingroup\let\do\@makeother\dospecials
  \catcode`\{=1 \catcode`\}=2 \doi@aux}
\providecommand{\doi@aux}[1]{\endgroup\texttt{#1}}
\makeatother
\providecommand*\mcitethebibliography{\thebibliography}
\csname @ifundefined\endcsname{endmcitethebibliography}
  {\let\endmcitethebibliography\endthebibliography}{}
\begin{mcitethebibliography}{82}
\providecommand*\natexlab[1]{#1}
\providecommand*\mciteSetBstSublistMode[1]{}
\providecommand*\mciteSetBstMaxWidthForm[2]{}
\providecommand*\mciteBstWouldAddEndPuncttrue
  {\def\EndOfBibitem{\unskip.}}
\providecommand*\mciteBstWouldAddEndPunctfalse
  {\let\EndOfBibitem\relax}
\providecommand*\mciteSetBstMidEndSepPunct[3]{}
\providecommand*\mciteSetBstSublistLabelBeginEnd[3]{}
\providecommand*\EndOfBibitem{}
\mciteSetBstSublistMode{f}
\mciteSetBstMaxWidthForm{subitem}{(\alph{mcitesubitemcount})}
\mciteSetBstSublistLabelBeginEnd
  {\mcitemaxwidthsubitemform\space}
  {\relax}
  {\relax}

\bibitem[Nielsen and Chuang(2000)Nielsen, and Chuang]{nielsen_chuang}
Nielsen,~M.~A.; Chuang,~I.~L. \emph{Quantum Computation and Quantum
  Information}; Cambridge University Press, 2000\relax
\mciteBstWouldAddEndPuncttrue
\mciteSetBstMidEndSepPunct{\mcitedefaultmidpunct}
{\mcitedefaultendpunct}{\mcitedefaultseppunct}\relax
\EndOfBibitem
\bibitem[Cao \latin{et~al.}(2019)Cao, Romero, Olson, Degroote, Johnson,
  Kieferov{\'{a}}, Kivlichan, Menke, Peropadre, Sawaya, Sim, Veis, and
  Aspuru-Guzik]{Cao2019}
Cao,~Y.; Romero,~J.; Olson,~J.~P.; Degroote,~M.; Johnson,~P.~D.;
  Kieferov{\'{a}},~M.; Kivlichan,~I.~D.; Menke,~T.; Peropadre,~B.; Sawaya,~N.
  P.~D.; Sim,~S.; Veis,~L.; Aspuru-Guzik,~A. Quantum Chemistry in the Age of
  Quantum Computing. \emph{Chem. Rev.} \textbf{2019}, \emph{119},
  10856--10915\relax
\mciteBstWouldAddEndPuncttrue
\mciteSetBstMidEndSepPunct{\mcitedefaultmidpunct}
{\mcitedefaultendpunct}{\mcitedefaultseppunct}\relax
\EndOfBibitem
\bibitem[McArdle \latin{et~al.}(2020)McArdle, Endo, Aspuru-Guzik, Benjamin, and
  Yuan]{McArdle2020}
McArdle,~S.; Endo,~S.; Aspuru-Guzik,~A.; Benjamin,~S.~C.; Yuan,~X. Quantum
  computational chemistry. \emph{Rev. Mod. Phys.} \textbf{2020},
  \emph{92}\relax
\mciteBstWouldAddEndPuncttrue
\mciteSetBstMidEndSepPunct{\mcitedefaultmidpunct}
{\mcitedefaultendpunct}{\mcitedefaultseppunct}\relax
\EndOfBibitem
\bibitem[Head-Marsden \latin{et~al.}(2020)Head-Marsden, Flick, Ciccarino, and
  Narang]{HeadMarsden2020}
Head-Marsden,~K.; Flick,~J.; Ciccarino,~C.~J.; Narang,~P. Quantum Information
  and Algorithms for Correlated Quantum Matter. \emph{Chem. Rev.}
  \textbf{2020}, \emph{121}, 3061--3120\relax
\mciteBstWouldAddEndPuncttrue
\mciteSetBstMidEndSepPunct{\mcitedefaultmidpunct}
{\mcitedefaultendpunct}{\mcitedefaultseppunct}\relax
\EndOfBibitem
\bibitem[Bauer \latin{et~al.}(2020)Bauer, Bravyi, Motta, and Chan]{Bauer2020}
Bauer,~B.; Bravyi,~S.; Motta,~M.; Chan,~G. K.-L. Quantum Algorithms for Quantum
  Chemistry and Quantum Materials Science. \emph{Chem. Rev.} \textbf{2020},
  \emph{120}, 12685--12717\relax
\mciteBstWouldAddEndPuncttrue
\mciteSetBstMidEndSepPunct{\mcitedefaultmidpunct}
{\mcitedefaultendpunct}{\mcitedefaultseppunct}\relax
\EndOfBibitem
\bibitem[Motta and Rice(2021)Motta, and Rice]{Motta2021}
Motta,~M.; Rice,~J.~E. Emerging quantum computing algorithms for quantum
  chemistry. \emph{{WIREs} Comput. Mol. Sci.} \textbf{2021}, \emph{12}\relax
\mciteBstWouldAddEndPuncttrue
\mciteSetBstMidEndSepPunct{\mcitedefaultmidpunct}
{\mcitedefaultendpunct}{\mcitedefaultseppunct}\relax
\EndOfBibitem
\bibitem[Aspuru-Guzik \latin{et~al.}(2005)Aspuru-Guzik, Dutoi, Love, and
  Head-Gordon]{aspuru-guzik_2005}
Aspuru-Guzik,~A.; Dutoi,~A.~D.; Love,~P.~J.; Head-Gordon,~M. Simulated Quantum
  Computation of Molecular Energies. \emph{Science} \textbf{2005}, \emph{309},
  1704--1707\relax
\mciteBstWouldAddEndPuncttrue
\mciteSetBstMidEndSepPunct{\mcitedefaultmidpunct}
{\mcitedefaultendpunct}{\mcitedefaultseppunct}\relax
\EndOfBibitem
\bibitem[Abrams and Lloyd(1997)Abrams, and Lloyd]{abrams_1997}
Abrams,~D.~S.; Lloyd,~S. Simulation of Many-Body Fermi Systems on a Universal
  Quantum Computer. \emph{Phys. Rev. Lett.} \textbf{1997}, \emph{79},
  2586--2589\relax
\mciteBstWouldAddEndPuncttrue
\mciteSetBstMidEndSepPunct{\mcitedefaultmidpunct}
{\mcitedefaultendpunct}{\mcitedefaultseppunct}\relax
\EndOfBibitem
\bibitem[Abrams and Lloyd(1999)Abrams, and Lloyd]{abrams_1999}
Abrams,~D.~S.; Lloyd,~S. A quantum algorithm providing exponential speed
  increase for finding eigenvalues and eigenvectors. \emph{Phys. Rev. Lett.}
  \textbf{1999}, \emph{83}, 5162--5165\relax
\mciteBstWouldAddEndPuncttrue
\mciteSetBstMidEndSepPunct{\mcitedefaultmidpunct}
{\mcitedefaultendpunct}{\mcitedefaultseppunct}\relax
\EndOfBibitem
\bibitem[Reiher \latin{et~al.}(2017)Reiher, Wiebe, Svore, Wecker, and
  Troyer]{Reiher2017}
Reiher,~M.; Wiebe,~N.; Svore,~K.~M.; Wecker,~D.; Troyer,~M. Elucidating
  reaction mechanisms on quantum computers. \emph{PNAS} \textbf{2017},
  \emph{114}, 7555--7560\relax
\mciteBstWouldAddEndPuncttrue
\mciteSetBstMidEndSepPunct{\mcitedefaultmidpunct}
{\mcitedefaultendpunct}{\mcitedefaultseppunct}\relax
\EndOfBibitem
\bibitem[Terhal(2015)]{Terhal2015}
Terhal,~B.~M. Quantum error correction for quantum memories. \emph{Rev. Mod.
  Phys.} \textbf{2015}, \emph{87}, 307--346\relax
\mciteBstWouldAddEndPuncttrue
\mciteSetBstMidEndSepPunct{\mcitedefaultmidpunct}
{\mcitedefaultendpunct}{\mcitedefaultseppunct}\relax
\EndOfBibitem
\bibitem[Preskill(2018)]{Preskill2018}
Preskill,~J. Quantum Computing in the {NISQ} era and beyond. \emph{Quantum}
  \textbf{2018}, \emph{2}, 79\relax
\mciteBstWouldAddEndPuncttrue
\mciteSetBstMidEndSepPunct{\mcitedefaultmidpunct}
{\mcitedefaultendpunct}{\mcitedefaultseppunct}\relax
\EndOfBibitem
\bibitem[Peruzzo \latin{et~al.}(2014)Peruzzo, McClean, Shadbolt, Yung, Zhou,
  Love, Aspuru-Guzik, and O'Brien]{Peruzzo2014}
Peruzzo,~A.; McClean,~J.; Shadbolt,~P.; Yung,~M.-H.; Zhou,~X.-Q.; Love,~P.~J.;
  Aspuru-Guzik,~A.; O'Brien,~J.~L. A variational eigenvalue solver on a
  photonic quantum processor. \emph{Nat. Commun.} \textbf{2014}, \emph{5}\relax
\mciteBstWouldAddEndPuncttrue
\mciteSetBstMidEndSepPunct{\mcitedefaultmidpunct}
{\mcitedefaultendpunct}{\mcitedefaultseppunct}\relax
\EndOfBibitem
\bibitem[McClean \latin{et~al.}(2016)McClean, Romero, Babbush, and
  Aspuru-Guzik]{McClean2016}
McClean,~J.~R.; Romero,~J.; Babbush,~R.; Aspuru-Guzik,~A. The theory of
  variational hybrid quantum-classical algorithms. \emph{New J. Phys.}
  \textbf{2016}, \emph{18}, 023023\relax
\mciteBstWouldAddEndPuncttrue
\mciteSetBstMidEndSepPunct{\mcitedefaultmidpunct}
{\mcitedefaultendpunct}{\mcitedefaultseppunct}\relax
\EndOfBibitem
\bibitem[Tilly \latin{et~al.}(2022)Tilly, Chen, Cao, Picozzi, Setia, Li, Grant,
  Wossnig, Rungger, Booth, and Tennyson]{Tilly2022}
Tilly,~J.; Chen,~H.; Cao,~S.; Picozzi,~D.; Setia,~K.; Li,~Y.; Grant,~E.;
  Wossnig,~L.; Rungger,~I.; Booth,~G.~H.; Tennyson,~J. The Variational Quantum
  Eigensolver: A review of methods and best practices. \emph{Phys. Rep.}
  \textbf{2022}, \emph{986}, 1--128\relax
\mciteBstWouldAddEndPuncttrue
\mciteSetBstMidEndSepPunct{\mcitedefaultmidpunct}
{\mcitedefaultendpunct}{\mcitedefaultseppunct}\relax
\EndOfBibitem
\bibitem[Romero \latin{et~al.}(2018)Romero, Babbush, McClean, Hempel, Love, and
  Aspuru-Guzik]{Romero2018}
Romero,~J.; Babbush,~R.; McClean,~J.~R.; Hempel,~C.; Love,~P.~J.;
  Aspuru-Guzik,~A. Strategies for quantum computing molecular energies using
  the unitary coupled cluster ansatz. \emph{Quantum Sci. Technol.}
  \textbf{2018}, \emph{4}, 014008\relax
\mciteBstWouldAddEndPuncttrue
\mciteSetBstMidEndSepPunct{\mcitedefaultmidpunct}
{\mcitedefaultendpunct}{\mcitedefaultseppunct}\relax
\EndOfBibitem
\bibitem[Smart and Mazziotti(2021)Smart, and Mazziotti]{Smart2021}
Smart,~S.~E.; Mazziotti,~D.~A. Quantum Solver of Contracted Eigenvalue
  Equations for Scalable Molecular Simulations on Quantum Computing Devices.
  \emph{Phys. Rev. Lett.} \textbf{2021}, \emph{126}\relax
\mciteBstWouldAddEndPuncttrue
\mciteSetBstMidEndSepPunct{\mcitedefaultmidpunct}
{\mcitedefaultendpunct}{\mcitedefaultseppunct}\relax
\EndOfBibitem
\bibitem[Stair and Evangelista(2021)Stair, and Evangelista]{Stair2021}
Stair,~N.~H.; Evangelista,~F.~A. Simulating Many-Body Systems with a Projective
  Quantum Eigensolver. \emph{{PRX} Quantum} \textbf{2021}, \emph{2}\relax
\mciteBstWouldAddEndPuncttrue
\mciteSetBstMidEndSepPunct{\mcitedefaultmidpunct}
{\mcitedefaultendpunct}{\mcitedefaultseppunct}\relax
\EndOfBibitem
\bibitem[Motta \latin{et~al.}(2020)Motta, Gujarati, Rice, Kumar, Masteran,
  Latone, Lee, Valeev, and Takeshita]{Motta2020}
Motta,~M.; Gujarati,~T.~P.; Rice,~J.~E.; Kumar,~A.; Masteran,~C.;
  Latone,~J.~A.; Lee,~E.; Valeev,~E.~F.; Takeshita,~T.~Y. Quantum simulation of
  electronic structure with a transcorrelated Hamiltonian: improved accuracy
  with a smaller footprint on the quantum computer. \emph{Phys. Chem. Chem.
  Phys.} \textbf{2020}, \emph{22}, 24270--24281\relax
\mciteBstWouldAddEndPuncttrue
\mciteSetBstMidEndSepPunct{\mcitedefaultmidpunct}
{\mcitedefaultendpunct}{\mcitedefaultseppunct}\relax
\EndOfBibitem
\bibitem[Sokolov \latin{et~al.}(2023)Sokolov, Dobrautz, Luo, Alavi, and
  Tavernelli]{Sokolov2023}
Sokolov,~I.~O.; Dobrautz,~W.; Luo,~H.; Alavi,~A.; Tavernelli,~I. Orders of
  magnitude increased accuracy for quantum many-body problems on quantum
  computers via an exact transcorrelated method. \emph{Phys. Rev. Res.}
  \textbf{2023}, \emph{5}\relax
\mciteBstWouldAddEndPuncttrue
\mciteSetBstMidEndSepPunct{\mcitedefaultmidpunct}
{\mcitedefaultendpunct}{\mcitedefaultseppunct}\relax
\EndOfBibitem
\bibitem[Bauman \latin{et~al.}(2019)Bauman, Bylaska, Krishnamoorthy, Low,
  Wiebe, Granade, Roetteler, Troyer, and Kowalski]{Bauman2019}
Bauman,~N.~P.; Bylaska,~E.~J.; Krishnamoorthy,~S.; Low,~G.~H.; Wiebe,~N.;
  Granade,~C.~E.; Roetteler,~M.; Troyer,~M.; Kowalski,~K. Downfolding of
  many-body Hamiltonians using active-space models: Extension of the sub-system
  embedding sub-algebras approach to unitary coupled cluster formalisms.
  \emph{J. Chem. Phys.} \textbf{2019}, \emph{151}\relax
\mciteBstWouldAddEndPuncttrue
\mciteSetBstMidEndSepPunct{\mcitedefaultmidpunct}
{\mcitedefaultendpunct}{\mcitedefaultseppunct}\relax
\EndOfBibitem
\bibitem[Bauman \latin{et~al.}(2021)Bauman, Chl{\'{a}}dek, Veis, Pittner, and
  Kowalski]{Bauman2021}
Bauman,~N.~P.; Chl{\'{a}}dek,~J.; Veis,~L.; Pittner,~J.; Kowalski,~K.
  Variational quantum eigensolver for approximate diagonalization of downfolded
  Hamiltonians using generalized unitary coupled cluster ansatz. \emph{Quantum
  Sci. Technol.} \textbf{2021}, \emph{6}, 034008\relax
\mciteBstWouldAddEndPuncttrue
\mciteSetBstMidEndSepPunct{\mcitedefaultmidpunct}
{\mcitedefaultendpunct}{\mcitedefaultseppunct}\relax
\EndOfBibitem
\bibitem[Le and Tran(2023)Le, and Tran]{Le2023}
Le,~N.~T.; Tran,~L.~N. Correlated Reference-Assisted Variational Quantum
  Eigensolver. \emph{J. Phys. Chem. A} \textbf{2023}, \emph{127},
  5222--5230\relax
\mciteBstWouldAddEndPuncttrue
\mciteSetBstMidEndSepPunct{\mcitedefaultmidpunct}
{\mcitedefaultendpunct}{\mcitedefaultseppunct}\relax
\EndOfBibitem
\bibitem[McClean \latin{et~al.}(2017)McClean, Kimchi-Schwartz, Carter, and
  de~Jong]{McClean2017}
McClean,~J.~R.; Kimchi-Schwartz,~M.~E.; Carter,~J.; de~Jong,~W.~A. Hybrid
  quantum-classical hierarchy for mitigation of decoherence and determination
  of excited states. \emph{Phys. Rev. A} \textbf{2017}, \emph{95}\relax
\mciteBstWouldAddEndPuncttrue
\mciteSetBstMidEndSepPunct{\mcitedefaultmidpunct}
{\mcitedefaultendpunct}{\mcitedefaultseppunct}\relax
\EndOfBibitem
\bibitem[Takeshita \latin{et~al.}(2020)Takeshita, Rubin, Jiang, Lee, Babbush,
  and McClean]{Takeshita2020}
Takeshita,~T.; Rubin,~N.~C.; Jiang,~Z.; Lee,~E.; Babbush,~R.; McClean,~J.~R.
  Increasing the Representation Accuracy of Quantum Simulations of Chemistry
  without Extra Quantum Resources. \emph{Phys. Rev. X} \textbf{2020},
  \emph{10}\relax
\mciteBstWouldAddEndPuncttrue
\mciteSetBstMidEndSepPunct{\mcitedefaultmidpunct}
{\mcitedefaultendpunct}{\mcitedefaultseppunct}\relax
\EndOfBibitem
\bibitem[Krompiec and Ramo(2022)Krompiec, and Ramo]{Krompiec2022}
Krompiec,~M.; Ramo,~D.~M. Strongly Contracted N-Electron Valence State
  Perturbation Theory Using Reduced Density Matrices from a Quantum Computer.
  2022\relax
\mciteBstWouldAddEndPuncttrue
\mciteSetBstMidEndSepPunct{\mcitedefaultmidpunct}
{\mcitedefaultendpunct}{\mcitedefaultseppunct}\relax
\EndOfBibitem
\bibitem[Tammaro \latin{et~al.}(2023)Tammaro, Galli, Rice, and
  Motta]{Tammaro2023}
Tammaro,~A.; Galli,~D.~E.; Rice,~J.~E.; Motta,~M. N-Electron Valence
  Perturbation Theory with Reference Wave Functions from Quantum Computing:
  Application to the Relative Stability of Hydroxide Anion and Hydroxyl
  Radical. \emph{J. Phys. Chem. A} \textbf{2023}, \emph{127}, 817--827\relax
\mciteBstWouldAddEndPuncttrue
\mciteSetBstMidEndSepPunct{\mcitedefaultmidpunct}
{\mcitedefaultendpunct}{\mcitedefaultseppunct}\relax
\EndOfBibitem
\bibitem[Pernal(2018)]{Pernal2018}
Pernal,~K. Electron Correlation from the Adiabatic Connection for
  Multireference Wave Functions. \emph{Phys. Rev. Lett.} \textbf{2018},
  \emph{120}\relax
\mciteBstWouldAddEndPuncttrue
\mciteSetBstMidEndSepPunct{\mcitedefaultmidpunct}
{\mcitedefaultendpunct}{\mcitedefaultseppunct}\relax
\EndOfBibitem
\bibitem[Pastorczak and Pernal(2018)Pastorczak, and Pernal]{Pastorczak2018}
Pastorczak,~E.; Pernal,~K. Correlation Energy from the Adiabatic Connection
  Formalism for Complete Active Space Wave Functions. \emph{J. Chem. Theory
  Comput.} \textbf{2018}, \emph{14}, 3493--3503\relax
\mciteBstWouldAddEndPuncttrue
\mciteSetBstMidEndSepPunct{\mcitedefaultmidpunct}
{\mcitedefaultendpunct}{\mcitedefaultseppunct}\relax
\EndOfBibitem
\bibitem[Pastorczak \latin{et~al.}(2019)Pastorczak, Hapka, Veis, and
  Pernal]{Pastorczak2019}
Pastorczak,~E.; Hapka,~M.; Veis,~L.; Pernal,~K. Capturing the Dynamic
  Correlation for Arbitrary Spin-Symmetry {CASSCF} Reference with Adiabatic
  Connection Approaches: Insights into the Electronic Structure of the
  Tetramethyleneethane Diradical. \emph{J. Phys. Chem. Lett.} \textbf{2019},
  \emph{10}, 4668--4674\relax
\mciteBstWouldAddEndPuncttrue
\mciteSetBstMidEndSepPunct{\mcitedefaultmidpunct}
{\mcitedefaultendpunct}{\mcitedefaultseppunct}\relax
\EndOfBibitem
\bibitem[Drwal \latin{et~al.}(2022)Drwal, Beran, Hapka, Modrzejewski,
  Sok{\'{o}}{\l}, Veis, and Pernal]{Drwal2022}
Drwal,~D.; Beran,~P.; Hapka,~M.; Modrzejewski,~M.; Sok{\'{o}}{\l},~A.;
  Veis,~L.; Pernal,~K. Efficient Adiabatic Connection Approach for Strongly
  Correlated Systems: Application to Singlet{\textendash}Triplet Gaps of
  Biradicals. \emph{J. Phys. Chem. Lett.} \textbf{2022}, \emph{13},
  4570--4578\relax
\mciteBstWouldAddEndPuncttrue
\mciteSetBstMidEndSepPunct{\mcitedefaultmidpunct}
{\mcitedefaultendpunct}{\mcitedefaultseppunct}\relax
\EndOfBibitem
\bibitem[Boyn \latin{et~al.}(2021)Boyn, Lykhin, Smart, Gagliardi, and
  Mazziotti]{Boyn2021}
Boyn,~J.-N.; Lykhin,~A.~O.; Smart,~S.~E.; Gagliardi,~L.; Mazziotti,~D.~A.
  Quantum-classical hybrid algorithm for the simulation of all-electron
  correlation. \emph{J. Chem. Phys.} \textbf{2021}, \emph{155}\relax
\mciteBstWouldAddEndPuncttrue
\mciteSetBstMidEndSepPunct{\mcitedefaultmidpunct}
{\mcitedefaultendpunct}{\mcitedefaultseppunct}\relax
\EndOfBibitem
\bibitem[Veis \latin{et~al.}(2016)Veis, Vi{\v{s}}{\v{n}}{\'{a}}k, Nishizawa,
  Nakai, and Pittner]{Veis2016}
Veis,~L.; Vi{\v{s}}{\v{n}}{\'{a}}k,~J.; Nishizawa,~H.; Nakai,~H.; Pittner,~J.
  Quantum chemistry beyond Born-Oppenheimer approximation on a quantum
  computer: A simulated phase estimation study. \emph{Int. J. Quant. Chem.}
  \textbf{2016}, \emph{116}, 1328--1336\relax
\mciteBstWouldAddEndPuncttrue
\mciteSetBstMidEndSepPunct{\mcitedefaultmidpunct}
{\mcitedefaultendpunct}{\mcitedefaultseppunct}\relax
\EndOfBibitem
\bibitem[Pavošević \latin{et~al.}(2020)Pavošević, Culpitt, and
  Hammes-Schiffer]{pavosevic2020chemrev}
Pavošević,~F.; Culpitt,~T.; Hammes-Schiffer,~S. Multicomponent Quantum
  Chemistry: Integrating Electronic and Nuclear Quantum Effects via the
  Nuclear--Electronic Orbital Method. \emph{Chem. Rev.} \textbf{2020},
  \emph{120}, 4222--4253\relax
\mciteBstWouldAddEndPuncttrue
\mciteSetBstMidEndSepPunct{\mcitedefaultmidpunct}
{\mcitedefaultendpunct}{\mcitedefaultseppunct}\relax
\EndOfBibitem
\bibitem[Pavo{\v{s}}evi{\'{c}} and Hammes-Schiffer(2021)Pavo{\v{s}}evi{\'{c}},
  and Hammes-Schiffer]{Pavosevic2021}
Pavo{\v{s}}evi{\'{c}},~F.; Hammes-Schiffer,~S. Multicomponent Unitary Coupled
  Cluster and Equation-of-Motion for Quantum Computation. \emph{J. Chem. Theory
  Comput.} \textbf{2021}, \emph{17}, 3252--3258\relax
\mciteBstWouldAddEndPuncttrue
\mciteSetBstMidEndSepPunct{\mcitedefaultmidpunct}
{\mcitedefaultendpunct}{\mcitedefaultseppunct}\relax
\EndOfBibitem
\bibitem[Szabo and Ostlund(1996)Szabo, and Ostlund]{szabo_ostlund}
Szabo,~A.; Ostlund,~N. \emph{Modern Quantum Chemistry: Introduction to Advanced
  Electronic Structure Theory}; Dover Publications, 1996\relax
\mciteBstWouldAddEndPuncttrue
\mciteSetBstMidEndSepPunct{\mcitedefaultmidpunct}
{\mcitedefaultendpunct}{\mcitedefaultseppunct}\relax
\EndOfBibitem
\bibitem[Claudino(2022)]{Claudino2022}
Claudino,~D. The basics of quantum computing for chemists. \emph{Int. J. Quant.
  Chem.} \textbf{2022}, \emph{122}\relax
\mciteBstWouldAddEndPuncttrue
\mciteSetBstMidEndSepPunct{\mcitedefaultmidpunct}
{\mcitedefaultendpunct}{\mcitedefaultseppunct}\relax
\EndOfBibitem
\bibitem[Jordan and Wigner(1928)Jordan, and Wigner]{jordan_1928}
Jordan,~P.; Wigner,~E. Uber das paulische Aquivalenzverbot. \emph{Z. Phys. A}
  \textbf{1928}, \emph{47}, 631\relax
\mciteBstWouldAddEndPuncttrue
\mciteSetBstMidEndSepPunct{\mcitedefaultmidpunct}
{\mcitedefaultendpunct}{\mcitedefaultseppunct}\relax
\EndOfBibitem
\bibitem[Whitfield \latin{et~al.}(2011)Whitfield, Biamonte, and
  Aspuru-Guzik]{whitfield_2010}
Whitfield,~J.~D.; Biamonte,~J.; Aspuru-Guzik,~A. Quantum Computing Resource
  Estimate of Molecular Energy Simulation. \emph{Mol. Phys.} \textbf{2011},
  \emph{109}, 735--750\relax
\mciteBstWouldAddEndPuncttrue
\mciteSetBstMidEndSepPunct{\mcitedefaultmidpunct}
{\mcitedefaultendpunct}{\mcitedefaultseppunct}\relax
\EndOfBibitem
\bibitem[Bravyi and Kitaev(2002)Bravyi, and Kitaev]{Bravyi2002}
Bravyi,~S.~B.; Kitaev,~A.~Y. Fermionic Quantum Computation. \emph{Ann. Phys.}
  \textbf{2002}, \emph{298}, 210--226\relax
\mciteBstWouldAddEndPuncttrue
\mciteSetBstMidEndSepPunct{\mcitedefaultmidpunct}
{\mcitedefaultendpunct}{\mcitedefaultseppunct}\relax
\EndOfBibitem
\bibitem[Seeley \latin{et~al.}(2012)Seeley, Richard, and Love]{Seeley2012}
Seeley,~J.~T.; Richard,~M.~J.; Love,~P.~J. The Bravyi-Kitaev transformation for
  quantum computation of electronic structure. \emph{J. Chem. Phys.}
  \textbf{2012}, \emph{137}, 224109\relax
\mciteBstWouldAddEndPuncttrue
\mciteSetBstMidEndSepPunct{\mcitedefaultmidpunct}
{\mcitedefaultendpunct}{\mcitedefaultseppunct}\relax
\EndOfBibitem
\bibitem[Kandala \latin{et~al.}(2017)Kandala, Mezzacapo, Temme, Takita, Brink,
  Chow, and Gambetta]{Kandala2017}
Kandala,~A.; Mezzacapo,~A.; Temme,~K.; Takita,~M.; Brink,~M.; Chow,~J.~M.;
  Gambetta,~J.~M. Hardware-efficient variational quantum eigensolver for small
  molecules and quantum magnets. \emph{Nature} \textbf{2017}, \emph{549},
  242--246\relax
\mciteBstWouldAddEndPuncttrue
\mciteSetBstMidEndSepPunct{\mcitedefaultmidpunct}
{\mcitedefaultendpunct}{\mcitedefaultseppunct}\relax
\EndOfBibitem
\bibitem[Anand \latin{et~al.}(2022)Anand, Schleich, Alperin-Lea, Jensen, Sim,
  D{\'{\i}}az-Tinoco, Kottmann, Degroote, Izmaylov, and
  Aspuru-Guzik]{Anand2022}
Anand,~A.; Schleich,~P.; Alperin-Lea,~S.; Jensen,~P. W.~K.; Sim,~S.;
  D{\'{\i}}az-Tinoco,~M.; Kottmann,~J.~S.; Degroote,~M.; Izmaylov,~A.~F.;
  Aspuru-Guzik,~A. A quantum computing view on unitary coupled cluster theory.
  \emph{Chem. Soc. Rev.} \textbf{2022}, \emph{51}, 1659--1684\relax
\mciteBstWouldAddEndPuncttrue
\mciteSetBstMidEndSepPunct{\mcitedefaultmidpunct}
{\mcitedefaultendpunct}{\mcitedefaultseppunct}\relax
\EndOfBibitem
\bibitem[Lee \latin{et~al.}(2018)Lee, Huggins, Head-Gordon, and
  Whaley]{Lee2018}
Lee,~J.; Huggins,~W.~J.; Head-Gordon,~M.; Whaley,~K.~B. Generalized Unitary
  Coupled Cluster Wave functions for Quantum Computation. \emph{J. Chem. Theory
  Comput.} \textbf{2018}, \emph{15}, 311--324\relax
\mciteBstWouldAddEndPuncttrue
\mciteSetBstMidEndSepPunct{\mcitedefaultmidpunct}
{\mcitedefaultendpunct}{\mcitedefaultseppunct}\relax
\EndOfBibitem
\bibitem[Matsuzawa and Kurashige(2020)Matsuzawa, and Kurashige]{Matsuzawa2020}
Matsuzawa,~Y.; Kurashige,~Y. Jastrow-type Decomposition in Quantum Chemistry
  for Low-Depth Quantum Circuits. \emph{J. Chem. Theory Comput.} \textbf{2020},
  \emph{16}, 944--952\relax
\mciteBstWouldAddEndPuncttrue
\mciteSetBstMidEndSepPunct{\mcitedefaultmidpunct}
{\mcitedefaultendpunct}{\mcitedefaultseppunct}\relax
\EndOfBibitem
\bibitem[Culpitt \latin{et~al.}(2023)Culpitt, Tellgren, and
  Pavošević]{culpitt2023unitary}
Culpitt,~T.; Tellgren,~E.~I.; Pavošević,~F. Unitary Coupled-Cluster for
  Quantum Computation of Molecular Properties in a Strong Magnetic Field.
  \emph{arXiv preprint arXiv:2309.12240} \textbf{2023}, \relax
\mciteBstWouldAddEndPunctfalse
\mciteSetBstMidEndSepPunct{\mcitedefaultmidpunct}
{}{\mcitedefaultseppunct}\relax
\EndOfBibitem
\bibitem[Sokolov \latin{et~al.}(2020)Sokolov, Barkoutsos, Ollitrault,
  Greenberg, Rice, Pistoia, and Tavernelli]{Sokolov2020}
Sokolov,~I.~O.; Barkoutsos,~P.~K.; Ollitrault,~P.~J.; Greenberg,~D.; Rice,~J.;
  Pistoia,~M.; Tavernelli,~I. Quantum orbital-optimized unitary coupled cluster
  methods in the strongly correlated regime: Can quantum algorithms outperform
  their classical equivalents? \emph{J. Chem. Phys.} \textbf{2020}, \emph{152},
  124107\relax
\mciteBstWouldAddEndPuncttrue
\mciteSetBstMidEndSepPunct{\mcitedefaultmidpunct}
{\mcitedefaultendpunct}{\mcitedefaultseppunct}\relax
\EndOfBibitem
\bibitem[Evangelista \latin{et~al.}(2019)Evangelista, Chan, and
  Scuseria]{Evangelista2019}
Evangelista,~F.~A.; Chan,~G. K.-L.; Scuseria,~G.~E. Exact parameterization of
  fermionic wave functions via unitary coupled cluster theory. \emph{J. Chem.
  Phys.} \textbf{2019}, \emph{151}, 244112\relax
\mciteBstWouldAddEndPuncttrue
\mciteSetBstMidEndSepPunct{\mcitedefaultmidpunct}
{\mcitedefaultendpunct}{\mcitedefaultseppunct}\relax
\EndOfBibitem
\bibitem[Grimsley \latin{et~al.}(2019)Grimsley, Economou, Barnes, and
  Mayhall]{Grimsley2019}
Grimsley,~H.~R.; Economou,~S.~E.; Barnes,~E.; Mayhall,~N.~J. An adaptive
  variational algorithm for exact molecular simulations on a quantum computer.
  \emph{Nat. Commun.} \textbf{2019}, \emph{10}\relax
\mciteBstWouldAddEndPuncttrue
\mciteSetBstMidEndSepPunct{\mcitedefaultmidpunct}
{\mcitedefaultendpunct}{\mcitedefaultseppunct}\relax
\EndOfBibitem
\bibitem[Ryabinkin \latin{et~al.}(2018)Ryabinkin, Yen, Genin, and
  Izmaylov]{Ryabinkin2018}
Ryabinkin,~I.~G.; Yen,~T.-C.; Genin,~S.~N.; Izmaylov,~A.~F. Qubit Coupled
  Cluster Method: A Systematic Approach to Quantum Chemistry on a Quantum
  Computer. \emph{J. Chem. Theory Comput.} \textbf{2018}, \emph{14},
  6317--6326\relax
\mciteBstWouldAddEndPuncttrue
\mciteSetBstMidEndSepPunct{\mcitedefaultmidpunct}
{\mcitedefaultendpunct}{\mcitedefaultseppunct}\relax
\EndOfBibitem
\bibitem[Tang \latin{et~al.}(2021)Tang, Shkolnikov, Barron, Grimsley, Mayhall,
  Barnes, and Economou]{Tang2021}
Tang,~H.~L.; Shkolnikov,~V.; Barron,~G.~S.; Grimsley,~H.~R.; Mayhall,~N.~J.;
  Barnes,~E.; Economou,~S.~E. Qubit-{ADAPT}-{VQE}: An Adaptive Algorithm for
  Constructing Hardware-Efficient Ans\"{a}tze on a Quantum Processor.
  \emph{{PRX} Quantum} \textbf{2021}, \emph{2}\relax
\mciteBstWouldAddEndPuncttrue
\mciteSetBstMidEndSepPunct{\mcitedefaultmidpunct}
{\mcitedefaultendpunct}{\mcitedefaultseppunct}\relax
\EndOfBibitem
\bibitem[Matou{\v{s}}ek \latin{et~al.}(2023)Matou{\v{s}}ek, Hapka, Veis, and
  Pernal]{Matousek2023}
Matou{\v{s}}ek,~M.; Hapka,~M.; Veis,~L.; Pernal,~K. Toward more accurate
  adiabatic connection approach for multireference wavefunctions. \emph{J.
  Chem. Phys.} \textbf{2023}, \emph{158}\relax
\mciteBstWouldAddEndPuncttrue
\mciteSetBstMidEndSepPunct{\mcitedefaultmidpunct}
{\mcitedefaultendpunct}{\mcitedefaultseppunct}\relax
\EndOfBibitem
\bibitem[Chatterjee and Pernal(2012)Chatterjee, and Pernal]{Chatterjee2012}
Chatterjee,~K.; Pernal,~K. Excitation energies from extended random phase
  approximation employed with approximate one- and two-electron reduced density
  matrices. \emph{J. Chem. Phys.} \textbf{2012}, \emph{137}\relax
\mciteBstWouldAddEndPuncttrue
\mciteSetBstMidEndSepPunct{\mcitedefaultmidpunct}
{\mcitedefaultendpunct}{\mcitedefaultseppunct}\relax
\EndOfBibitem
\bibitem[Rowe(1968)]{Rowe1968}
Rowe,~D.~J. Equations-of-Motion Method and the Extended Shell Model. \emph{Rev.
  Mod. Phys.} \textbf{1968}, \emph{40}, 153--166\relax
\mciteBstWouldAddEndPuncttrue
\mciteSetBstMidEndSepPunct{\mcitedefaultmidpunct}
{\mcitedefaultendpunct}{\mcitedefaultseppunct}\relax
\EndOfBibitem
\bibitem[Ollitrault \latin{et~al.}(2020)Ollitrault, Kandala, Chen, Barkoutsos,
  Mezzacapo, Pistoia, Sheldon, Woerner, Gambetta, and
  Tavernelli]{Ollitrault2020}
Ollitrault,~P.~J.; Kandala,~A.; Chen,~C.-F.; Barkoutsos,~P.~K.; Mezzacapo,~A.;
  Pistoia,~M.; Sheldon,~S.; Woerner,~S.; Gambetta,~J.~M.; Tavernelli,~I.
  Quantum equation of motion for computing molecular excitation energies on a
  noisy quantum processor. \emph{Phys. Rev. Res.} \textbf{2020}, \emph{2}\relax
\mciteBstWouldAddEndPuncttrue
\mciteSetBstMidEndSepPunct{\mcitedefaultmidpunct}
{\mcitedefaultendpunct}{\mcitedefaultseppunct}\relax
\EndOfBibitem
\bibitem[Asthana \latin{et~al.}(2023)Asthana, Kumar, Abraham, Grimsley, Zhang,
  Cincio, Tretiak, Dub, Economou, Barnes, and Mayhall]{Asthana2023}
Asthana,~A.; Kumar,~A.; Abraham,~V.; Grimsley,~H.; Zhang,~Y.; Cincio,~L.;
  Tretiak,~S.; Dub,~P.~A.; Economou,~S.~E.; Barnes,~E.; Mayhall,~N.~J. Quantum
  self-consistent equation-of-motion method for computing molecular excitation
  energies, ionization potentials, and electron affinities on a quantum
  computer. \emph{Chem. Sci.} \textbf{2023}, \emph{14}, 2405--2418\relax
\mciteBstWouldAddEndPuncttrue
\mciteSetBstMidEndSepPunct{\mcitedefaultmidpunct}
{\mcitedefaultendpunct}{\mcitedefaultseppunct}\relax
\EndOfBibitem
\bibitem[Pavošević \latin{et~al.}(2023)Pavošević, Tavernelli, and
  Rubio]{pavosevic2023spinflip}
Pavošević,~F.; Tavernelli,~I.; Rubio,~A. Spin-Flip Unitary Coupled Cluster
  Method: Toward Accurate Description of Strong Electron Correlation on Quantum
  Computers. \emph{J. Phys. Chem. Lett.} \textbf{2023}, \emph{14},
  7876--7882\relax
\mciteBstWouldAddEndPuncttrue
\mciteSetBstMidEndSepPunct{\mcitedefaultmidpunct}
{\mcitedefaultendpunct}{\mcitedefaultseppunct}\relax
\EndOfBibitem
\bibitem[Beran \latin{et~al.}(2021)Beran, Matou{\v{s}}ek, Hapka, Pernal, and
  Veis]{Beran2021}
Beran,~P.; Matou{\v{s}}ek,~M.; Hapka,~M.; Pernal,~K.; Veis,~L. Density Matrix
  Renormalization Group with Dynamical Correlation via Adiabatic Connection.
  \emph{J. Chem. Theory Comput.} \textbf{2021}, \emph{17}, 7575--7585\relax
\mciteBstWouldAddEndPuncttrue
\mciteSetBstMidEndSepPunct{\mcitedefaultmidpunct}
{\mcitedefaultendpunct}{\mcitedefaultseppunct}\relax
\EndOfBibitem
\bibitem[Mizukami \latin{et~al.}(2020)Mizukami, Mitarai, Nakagawa, Yamamoto,
  Yan, and ya~Ohnishi]{Mizukami2020}
Mizukami,~W.; Mitarai,~K.; Nakagawa,~Y.~O.; Yamamoto,~T.; Yan,~T.;
  ya~Ohnishi,~Y. Orbital optimized unitary coupled cluster theory for quantum
  computer. \emph{Phys. Rev. Res.} \textbf{2020}, \emph{2}\relax
\mciteBstWouldAddEndPuncttrue
\mciteSetBstMidEndSepPunct{\mcitedefaultmidpunct}
{\mcitedefaultendpunct}{\mcitedefaultseppunct}\relax
\EndOfBibitem
\bibitem[Yalouz \latin{et~al.}(2021)Yalouz, Senjean, G\"{u}nther, Buda,
  O'Brien, and Visscher]{Yalouz2021}
Yalouz,~S.; Senjean,~B.; G\"{u}nther,~J.; Buda,~F.; O'Brien,~T.~E.;
  Visscher,~L. A state-averaged orbital-optimized hybrid
  quantum{\textendash}classical algorithm for a democratic description of
  ground and excited states. \emph{Quantum Sci. Technol.} \textbf{2021},
  \emph{6}, 024004\relax
\mciteBstWouldAddEndPuncttrue
\mciteSetBstMidEndSepPunct{\mcitedefaultmidpunct}
{\mcitedefaultendpunct}{\mcitedefaultseppunct}\relax
\EndOfBibitem
\bibitem[de~Gracia~Trivi{\~{n}}o \latin{et~al.}(2023)de~Gracia~Trivi{\~{n}}o,
  Delcey, and Wendin]{deGraciaTrivio2023}
de~Gracia~Trivi{\~{n}}o,~J.~A.; Delcey,~M.~G.; Wendin,~G. Complete Active Space
  Methods for {NISQ} Devices: The Importance of Canonical Orbital Optimization
  for Accuracy and Noise Resilience. \emph{J. Chem. Theory Comput.}
  \textbf{2023}, \emph{19}, 2863--2872\relax
\mciteBstWouldAddEndPuncttrue
\mciteSetBstMidEndSepPunct{\mcitedefaultmidpunct}
{\mcitedefaultendpunct}{\mcitedefaultseppunct}\relax
\EndOfBibitem
\bibitem[Fitzpatrick \latin{et~al.}(2022)Fitzpatrick, Nyk\"{a}nen, Talarico,
  Lunghi, Maniscalco, García-Pérez, and Knecht]{adapt_vqe_scf}
Fitzpatrick,~A.; Nyk\"{a}nen,~A.; Talarico,~N.~W.; Lunghi,~A.; Maniscalco,~S.;
  García-Pérez,~G.; Knecht,~S. A self-consistent field approach for the
  variational quantum eigensolver: orbital optimization goes adaptive. 2022;
  \url{https://arxiv.org/abs/2212.11405}\relax
\mciteBstWouldAddEndPuncttrue
\mciteSetBstMidEndSepPunct{\mcitedefaultmidpunct}
{\mcitedefaultendpunct}{\mcitedefaultseppunct}\relax
\EndOfBibitem
\bibitem[Helgaker \latin{et~al.}(2000)Helgaker, J{\o}rgensen, and
  Olsen]{olsen_bible}
Helgaker,~T.; J{\o}rgensen,~P.; Olsen,~J. \emph{Molecular Electronic Structure
  Theory}; John Wiley \& Sons, LTD: Chichester, 2000\relax
\mciteBstWouldAddEndPuncttrue
\mciteSetBstMidEndSepPunct{\mcitedefaultmidpunct}
{\mcitedefaultendpunct}{\mcitedefaultseppunct}\relax
\EndOfBibitem
\bibitem[Sun \latin{et~al.}(2017)Sun, Yang, and Chan]{Sun2017}
Sun,~Q.; Yang,~J.; Chan,~G. K.-L. A general second order complete active space
  self-consistent-field solver for large-scale systems. \emph{Chem. Phys.
  Lett.} \textbf{2017}, \emph{683}, 291--299\relax
\mciteBstWouldAddEndPuncttrue
\mciteSetBstMidEndSepPunct{\mcitedefaultmidpunct}
{\mcitedefaultendpunct}{\mcitedefaultseppunct}\relax
\EndOfBibitem
\bibitem[Levine \latin{et~al.}(2020)Levine, Hait, Tubman, Lehtola, Whaley, and
  Head-Gordon]{Levine2020}
Levine,~D.~S.; Hait,~D.; Tubman,~N.~M.; Lehtola,~S.; Whaley,~K.~B.;
  Head-Gordon,~M. {CASSCF} with Extremely Large Active Spaces Using the
  Adaptive Sampling Configuration Interaction Method. \emph{J. Chem. Theory
  Comput.} \textbf{2020}, \emph{16}, 2340--2354\relax
\mciteBstWouldAddEndPuncttrue
\mciteSetBstMidEndSepPunct{\mcitedefaultmidpunct}
{\mcitedefaultendpunct}{\mcitedefaultseppunct}\relax
\EndOfBibitem
\bibitem[Pozun \latin{et~al.}(2013)Pozun, Su, and Jordan]{Pozun2013}
Pozun,~Z.~D.; Su,~X.; Jordan,~K.~D. Establishing the Ground State of the
  Disjoint Diradical Tetramethyleneethane with Quantum Monte Carlo. \emph{J.
  Am. Chem. Soc.} \textbf{2013}, \emph{135}, 13862--13869\relax
\mciteBstWouldAddEndPuncttrue
\mciteSetBstMidEndSepPunct{\mcitedefaultmidpunct}
{\mcitedefaultendpunct}{\mcitedefaultseppunct}\relax
\EndOfBibitem
\bibitem[Dunning(1989)]{Dunning1989}
Dunning,~T.~H. Gaussian Basis Sets for use in Correlated Molecular
  Calculations. I. The Atoms Boron Through Neon and Hydrogen. \emph{J. Chem.
  Phys.} \textbf{1989}, \emph{90}, 1007--1023\relax
\mciteBstWouldAddEndPuncttrue
\mciteSetBstMidEndSepPunct{\mcitedefaultmidpunct}
{\mcitedefaultendpunct}{\mcitedefaultseppunct}\relax
\EndOfBibitem
\bibitem[Stein and Reiher(2016)Stein, and Reiher]{Stein2016}
Stein,~C.~J.; Reiher,~M. Automated Selection of Active Orbital Spaces. \emph{J.
  Chem. Theory Comput.} \textbf{2016}, \emph{12}, 1760--1771\relax
\mciteBstWouldAddEndPuncttrue
\mciteSetBstMidEndSepPunct{\mcitedefaultmidpunct}
{\mcitedefaultendpunct}{\mcitedefaultseppunct}\relax
\EndOfBibitem
\bibitem[Veis and Pittner(2010)Veis, and Pittner]{Veis2010}
Veis,~L.; Pittner,~J. Quantum computing applied to calculations of molecular
  energies: {CH}2 benchmark. \emph{J. Chem. Phys.} \textbf{2010}, \emph{133},
  194106\relax
\mciteBstWouldAddEndPuncttrue
\mciteSetBstMidEndSepPunct{\mcitedefaultmidpunct}
{\mcitedefaultendpunct}{\mcitedefaultseppunct}\relax
\EndOfBibitem
\bibitem[Veis \latin{et~al.}(2012)Veis, Vi{\v{s}}{\v{n}}{\'{a}}k, Fleig,
  Knecht, Saue, Visscher, and Pittner]{Veis2012}
Veis,~L.; Vi{\v{s}}{\v{n}}{\'{a}}k,~J.; Fleig,~T.; Knecht,~S.; Saue,~T.;
  Visscher,~L.; Pittner,~J. Relativistic quantum chemistry on quantum
  computers. \emph{Phys. Rev. A} \textbf{2012}, \emph{85}\relax
\mciteBstWouldAddEndPuncttrue
\mciteSetBstMidEndSepPunct{\mcitedefaultmidpunct}
{\mcitedefaultendpunct}{\mcitedefaultseppunct}\relax
\EndOfBibitem
\bibitem[Veis and Pittner(2014)Veis, and Pittner]{Veis2014}
Veis,~L.; Pittner,~J. Adiabatic state preparation study of methylene. \emph{J.
  Chem. Phys.} \textbf{2014}, \emph{140}, 214111\relax
\mciteBstWouldAddEndPuncttrue
\mciteSetBstMidEndSepPunct{\mcitedefaultmidpunct}
{\mcitedefaultendpunct}{\mcitedefaultseppunct}\relax
\EndOfBibitem
\bibitem[Neese(2012)]{orca}
Neese,~F. \emph{WIREs Comput. Mol. Sci.} \textbf{2012}, \emph{2}, 73--78\relax
\mciteBstWouldAddEndPuncttrue
\mciteSetBstMidEndSepPunct{\mcitedefaultmidpunct}
{\mcitedefaultendpunct}{\mcitedefaultseppunct}\relax
\EndOfBibitem
\bibitem[Pernal \latin{et~al.}(2022)Pernal, Hapka, Przybytek, Modrzejewski, and
  Sokół]{gammcor}
Pernal,~K.; Hapka,~M.; Przybytek,~M.; Modrzejewski,~M.; Sokół,~A. GammCor
  code. \url{https://github.com/pernalk/GAMMCOR}, 2022\relax
\mciteBstWouldAddEndPuncttrue
\mciteSetBstMidEndSepPunct{\mcitedefaultmidpunct}
{\mcitedefaultendpunct}{\mcitedefaultseppunct}\relax
\EndOfBibitem
\bibitem[Brabec \latin{et~al.}(2020)Brabec, Brandejs, Kowalski, Xantheas,
  \"{O}rs Legeza, and Veis]{Brabec2020}
Brabec,~J.; Brandejs,~J.; Kowalski,~K.; Xantheas,~S.; \"{O}rs Legeza; Veis,~L.
  Massively parallel quantum chemical density matrix renormalization group
  method. \emph{J. Comp. Chem.} \textbf{2020}, \emph{42}, 534--544\relax
\mciteBstWouldAddEndPuncttrue
\mciteSetBstMidEndSepPunct{\mcitedefaultmidpunct}
{\mcitedefaultendpunct}{\mcitedefaultseppunct}\relax
\EndOfBibitem
\bibitem[{Qiskit contributors}(2023)]{Qiskit}
{Qiskit contributors} Qiskit: An Open-source Framework for Quantum Computing.
  2023\relax
\mciteBstWouldAddEndPuncttrue
\mciteSetBstMidEndSepPunct{\mcitedefaultmidpunct}
{\mcitedefaultendpunct}{\mcitedefaultseppunct}\relax
\EndOfBibitem
\bibitem[Bravyi \latin{et~al.}(2017)Bravyi, Gambetta, Mezzacapo, and
  Temme]{Bravyi_2017}
Bravyi,~S.; Gambetta,~J.~M.; Mezzacapo,~A.; Temme,~K. Tapering off qubits to
  simulate fermionic Hamiltonians. 2017;
  \url{https://arxiv.org/abs/1701.08213}\relax
\mciteBstWouldAddEndPuncttrue
\mciteSetBstMidEndSepPunct{\mcitedefaultmidpunct}
{\mcitedefaultendpunct}{\mcitedefaultseppunct}\relax
\EndOfBibitem
\bibitem[Setia \latin{et~al.}(2020)Setia, Chen, Rice, Mezzacapo, Pistoia, and
  Whitfield]{Setia2020}
Setia,~K.; Chen,~R.; Rice,~J.~E.; Mezzacapo,~A.; Pistoia,~M.; Whitfield,~J.~D.
  Reducing Qubit Requirements for Quantum Simulations Using Molecular Point
  Group Symmetries. \emph{J. Chem. Theory Comput.} \textbf{2020}, \emph{16},
  6091--6097\relax
\mciteBstWouldAddEndPuncttrue
\mciteSetBstMidEndSepPunct{\mcitedefaultmidpunct}
{\mcitedefaultendpunct}{\mcitedefaultseppunct}\relax
\EndOfBibitem
\bibitem[Veis \latin{et~al.}(2018)Veis, Antal{\'{\i}}k, \"{O}rs Legeza, Alavi,
  and Pittner]{tme_paper}
Veis,~L.; Antal{\'{\i}}k,~A.; \"{O}rs Legeza; Alavi,~A.; Pittner,~J. The
  Intricate Case of Tetramethyleneethane: A Full Configuration Interaction
  Quantum Monte Carlo Benchmark and Multireference Coupled Cluster Studies.
  \emph{J. Chem. Theory Comput.} \textbf{2018}, \emph{14}, 2439--2445\relax
\mciteBstWouldAddEndPuncttrue
\mciteSetBstMidEndSepPunct{\mcitedefaultmidpunct}
{\mcitedefaultendpunct}{\mcitedefaultseppunct}\relax
\EndOfBibitem
\bibitem[Clifford \latin{et~al.}(1998)Clifford, Wenthold, Lineberger, Ellison,
  Wang, Grabowski, Vila, and Jordan]{Clifford1998}
Clifford,~E.~P.; Wenthold,~P.~G.; Lineberger,~W.~C.; Ellison,~G.~B.;
  Wang,~C.~X.; Grabowski,~J.~J.; Vila,~F.; Jordan,~K.~D. Properties of
  tetramethyleneethane ({TME}) as revealed by ion chemistry and ion
  photoelectron spectroscopy. \emph{Journal of the Chemical Society, Perkin
  Transactions 2} \textbf{1998}, 1015--1022\relax
\mciteBstWouldAddEndPuncttrue
\mciteSetBstMidEndSepPunct{\mcitedefaultmidpunct}
{\mcitedefaultendpunct}{\mcitedefaultseppunct}\relax
\EndOfBibitem
\bibitem[Beran \latin{et~al.}(2023)Beran, Pernal, Pavošević, and
  Veis]{beran2023projection}
Beran,~P.; Pernal,~K.; Pavošević,~F.; Veis,~L. Projection-Based Density
  Matrix Renormalization Group in Density Functional Theory Embedding. \emph{J.
  Phys. Chem. Lett.} \textbf{2023}, \emph{14}, 716--722\relax
\mciteBstWouldAddEndPuncttrue
\mciteSetBstMidEndSepPunct{\mcitedefaultmidpunct}
{\mcitedefaultendpunct}{\mcitedefaultseppunct}\relax
\EndOfBibitem
\bibitem[Rossmannek \latin{et~al.}(2023)Rossmannek, Pavošević, Rubio, and
  Tavernelli]{rossmannek2023quantum}
Rossmannek,~M.; Pavošević,~F.; Rubio,~A.; Tavernelli,~I. Quantum Embedding
  Method for the Simulation of Strongly Correlated Systems on Quantum
  Computers. \emph{J. Phys. Chem. Lett.} \textbf{2023}, \emph{14},
  3491--3497\relax
\mciteBstWouldAddEndPuncttrue
\mciteSetBstMidEndSepPunct{\mcitedefaultmidpunct}
{\mcitedefaultendpunct}{\mcitedefaultseppunct}\relax
\EndOfBibitem
\end{mcitethebibliography}

\end{document}